# RADIATION PROCESSES IN SOLAR ENERGETICS

V.P. Yefimov, A.N. Dovbnya, A.S. Abyzov

National Science Center "Kharkov Institute of Physics and Technology"

The ways of creating of nanoscale systems (conducting filaments) in the ordered structure of a doped semiconductor in the aspect of the modern problems of nanotechnology to improve solar cells. Nuclear methods and processes of interaction of relativistic particles – gamma quants, electrons, protons, neutrons, and fragments of heavy elements with atoms of matter are used, radiation processes of cluster formation and directed Coulomb explosion are applied for formation of the ordered conducting structures.

## РАДИАЦИОННЫЕ ПРОЦЕССЫ В СОЛНЕЧНОЙ ЭНЕРГЕТИКЕ

В.П. Ефимов, А.Н. Довбня, А.С. Абызов

Показаны пути создания наноразмерных систем (проводящих нитей) в упорядоченной структуре легированного полупроводника в аспекте современных задач нанотехнологии по усовершенствованию солнечных батарей. Использованы ядерно-физические методы и процессы взаимодействия релятивистских частиц – гамма-квантов, электронов, протонов, нейтронов и осколков ядер тяжелых элементов с атомами вещества; применены радиационные процессы образования кластеров и их направленного кулоновского взрыва.

# РАДИАЦИОННЫЕ ПРОЦЕССЫ В СОЛНЕЧНОЙ ЭНЕРГЕТИКЕ

В.П. Ефимов, А.Н. Довбня, А.С. Абызов

Научно-исследовательский комплекс "Ускоритель"
Национального научного центра "Харьковский физико-технический институт"

## ВВЕДЕНИЕ

Среди развиваемых в настоящее время способов синтеза различных наноразмерных структур особое место отводится ядерно-пучковым технологиям. Уникальные свойства таких структур могут привести к появлению нового поколения электронных и оптоэлектронных приборов. В монокристаллическом кремнии процесс аморфизации происходит при возбуждении его электронной подсистемы легкими и тяжелыми осколками делящихся ядер после фрагментации структуры полупроводника в результате облучения ее пучком ускоренных электронов (ионов). Изучению проблемы образования наноструктур быстрыми ионами в различных материалах уделяют большое внимание крупнейшие ускорительные центры мира: GANIL (Каен, Франция), UNILAC (Дармштадт, Германия), У-300 (Дубна, Россия), TASCC (Чок-Риверские ядерные лаборатории, Канада). Экспериментально ядерные треки зафиксированы в ограниченном классе материалов, в частности, в некоторых металлах и их сплавах, диэлектриках, полимерах и полупроводниках. Но в ряде практически важных металлов и сплавов, а также в кремнии, являющемся базовым материалом полупроводниковой электроники, пока не удалось сформировать трековую структуру. Результаты исследований свидетельствуют о важной роли не только ионизационных потерь в образовании дефектной структуры в кремнии вдоль траектории движения ионов или осколков деления ядер, но и геометрических размеров атомной структуры материала мишени [1, 2].

Фрагментация кристаллической структуры представляет интерес с точки зрения создания аморфной фазы в объеме кристаллической матрицы. До недавнего времени аморфная фаза была получена только на поверхности кристалла методами разложения силиконов в тлеющем разряде, распыления структуры кремния в ВЧ-разряде или химического осаждения из газовой фазы. Исследуются возможности создания аморфных фаз в объеме монокристаллической кремниевой матрицы для формирования внутренних электростатических полей и скрытых проводящих каналов для вывода носителей тока из объема c-Si-структуры полупроводника. Воздействие осколков ядер урана, делящихся высокоэнергетичными γ-квантами, вызывает дефектообразование в структуре полупроводника. При исследовании радиационной модификации c-Si-полупроводников на вторичном пучке частиц линейного ускорителя электронов определяется влияние γ-облучения и осколков деления ядер $^{238}$U на время жизни носителей тока и величину удельного сопротивления структуры на различных глубинах дефектообразования. В результате увеличения дозы облучения, флюенса пучка на урановую мишень и энергии γ-квантов до области гигантского резонанса фотоделения ядер урана и формирования направленного кулоновского взрыва накопленного заряда для повышения диффузии атомов примеси в аморфном состоянии структуры с пониженной температурой отжига полупроводника можно управлять плотностью проводящих разупорядоченных структур. Увеличение удельной проводимости по структурным нарушениям вдоль пробега осколков используется для вывода носителей тока с малой подвижностью из объема полупроводника. При этом монокристаллическая структура кремниевой матрицы является транспортной средой носителей заряда, обладающих большой подвижностью. При разработке радиационных технологий на основе фотоделения ядер урана создаются дополнительные возможности в совершенствовании кремниевых фотопреобразователей для солнечной энергетики [3]. Предложенный метод по изменению структуры полупроводниковых материалов и полученные экспериментальные результаты представляют научный интерес также и для специалистов в области микротехнологии.

Анализируются особенности структур перспективных устройств кремниевой фотовольтаики с фотолюминесценцией коротковолнового солнечного излучения в область их максимальной фоточувствительности, с радиальными p-n-переходами в микроструктурах (микропроволоках) вертикального исполнения и с проводящими квантовыми нитями в объеме монокристалла кремния. Повышенная удельная электрическая проводимость в планарных структурах создается атомами кластерных объединений кремния в аморфном состоянии с атомами легированной примеси.

Массивы квантовых нитей являются основным структурным фактором при создании фотоэлементов нового поколения с повышенной конверсионной эффективностью, радиационной стабильностью и высоким ресурсом эксплуатации в наземной и космической энергетике.

В связи с изменением климата, сокращением углеводородного сырья, финансовым и энергетическим кризисом резко возрос интерес к источникам возобновляемой энергии. Солнечные электростанции – это перспективное направление развития энергетики в XXI-м веке. В стремлении избавиться от энергетической зависимости в Европейском Союзе разработан крупнейший инвестиционный проект (DESERTEC) стоимостью ~ $ 555 млрд. Проект предусматривает строительство 100 больших солнечных электростанций в пустыне Сахара. Если проект осуществится, то генерированной энергии будет достаточно для удовлетворения 15-25 % потребностей Европы и стран Северной Африки. Создан консорциум из крупнейших компаний и банков Европы по финансированию этого грандиозного строительства на основе параболлических концентраторов солнечного излучения для нагревания теплоносителя (масла) до параметров, пригодных к использованию в турбогенераторах (термоэффект).

В области прямого преобразования солнечного излучения в электрическую энергию (фотоэффект) потребность в фотопреобразователях только частично решается за счет существующего набора фотоэлементов (ФЭ). Основные потери солнечного излучения при трансформации его в электрическую энергию определяются фундаментальными ограничениями фотоэффекта в области IR-излучения ($h\nu \leq E_g$) и в области коротковолнового излучения ($h\nu \gg E_g$), а также низкой эффективностью вывода фотогенерированных носителей заряда из объема c-Si-фотопреобразователя [4]. Солнечные батареи (СБ) нашли применение в космонавтике, где занимают доминирующее положение среди других источников автономного энергопитания. СБ снабжают электроэнергией аппаратуру спутников и системы жизнеобеспечения космических кораблей и станций, а также заряжают электрохимические аккумуляторы, используемые на теневых участках орбиты. Возрастание требований к бортовым системам космических аппаратов (КА) приводит к необходимости создания солнечных батарей, обладающих более высокими энергетическими и эксплуатационными характеристиками. Основные из них – повышение кпд СБ, увеличение ресурса их эксплуатации на геостационарных орбитах в условиях повышенной радиации и возможности их функционирования при высокой концентрации солнечного излучения. В космических системах до 80 % применяются фотопреобразователи из монокристаллического кремния, а остальные 20 % -из GaAs-структур.

В наземных условиях СБ используются в быту и для питания устройств автоматики, переносных радиостанций и радиоприёмников, для катодной антикоррозионной защиты нефте- и газопроводов. Коммерческие модули солнечных элементов на основе кремния проходят 3-й этап в своем развитии и в настоящее время требуется значительное повышение кпд ФЭ и снижение их себестоимости. В результате модернизации структуры ФЭ за счет текстуризации фронтальной поверхности, оптимизации конструкции контактной гребенки, повышающей фоточувствительность в фиолетовой части солнечного спектра, и встроенного электрического поля (BSF) вблизи тыльной поверхности кпд увеличился уже до 17 %. Эта величина эффективности ФЭ оставалась непревзойденной в течение пятнадцати лет. Эффективность отдельных лабораторных образцов даже достигала ~20 %. Основным препятствием для достижения более высоких значений кпд является объемная рекомбинация носителей заряда (НЗ). Структура с пассивированным эмиттером и локальной диффузией НЗ на тыльной стороне стала вершиной процесса эволюции кремниевых ФЭ и было достигнуто наивысшее значение кпд 24 % при воздушной массе AM 1.5 [4]. Это термодинамический предел кремниевых ФЭ с одним p-n-переходом в условиях одного солнца. Для изготовления наземных фотоэлектрических преобразователей наиболее пригодными считаются полупроводники Si, CdTe, GaAs, InP. Около 91% энергии падающего светового потока можно преобразовывать в электрический ток в кремниевых ФЭП при решении проблемы вывода НЗ из объема полупроводника. Низкая стоимость кремния является определяющей при выборе материала для ФЭ. Энергетические характеристики СБ определяются полупроводниковым материалом, конструктивными особенностями ФЭ и количеством их в СБ (см. табл. 1) [5]

Таблица 1. Параметры промышленных солнечных батарей

| Материал | КПД,% | $P$, Вт | | | | | |
| --- | --- | --- | --- | --- | --- | --- | --- |
| | | Необлученные | | Облученные флюенсом электронов 1 МэВ | | | |
| | | | | $3 \cdot 10^{14}$ э/см$^2$ | | $1 \cdot 10^{15}$ э/см$^2$ | |
| | | 28°C | 50°C | 28°C | 50°C | 28°C | 50°C |
| Si | 14,8 | 170,9 | 149,5 | 129,0 | 112,2 | 113,0 | 98,8 |
| GaAs | 18,5 | 218,1 | 208,2 | 188,1 | 179,6 | 166,8 | 159,3 |
| GaInAs–GaAs–Ge | 21,5 | 253,5 | 242,8 | 223,0 | 211,9 | 192,7 | 183,0 |

Большинство коммерческих спутников оснащены СБ с размерами 2.5-3.5 м$^2$, но часть из них оснащается СБ ~ 70 м$^2$.

В магнитосфере Земли основными факторами деградации фотоэлектрических батарей являются возникновение центров окраски, снижение времени жизни неосновных носителей заряда, ограничивающих величину токосъема с ФЭ, и их ресурс эксплуатации. Необходима высокая радиационная стойкость просветляющих покрытий, а также наличие в структуре ФЭ глубокозалегающего p-n-перехода и проводящих квантовых структур для увеличения кпд и радиационной стойкости энергетических систем, работающих в радиационных поясах Земли.

В качестве наиболее вероятных материалов для фотоэлектрических систем преобразования солнечного излучения в электрическую энергию рассматриваются кремний и арсенид галлия, причём в последнем случае речь идёт о гетерофотопреобразователях со структурой AlGaAs-GaAs. Фотоэлектрические преобразователи на основе соединения мышьяка с галлием имеют, как известно, более высокий теоретический кпд чем кремниевые ФЭП, так как ширина запрещённой зоны у GaAs в полупроводниковых преобразователях оптимальна (1,4 эВ). У кремниевых ФЭП этот показатель равен 1,1 эВ и кпд таких батарей незначительно превышает 15%.

Солнечные ФЭП [6, 7] на основе GaInP/AlGaAs-гетероструктур и родственных им соединений $A^3B^5$ обеспечивают существенное увеличение кпд, удельного энергосъема и радиационной стойкости космических СБ по сравнению с батареями на основе кремния. В условиях космоса значение кпд ФЭП с одним p-n переходом в GaAs достигает величины ~25%, а в каскадных ФЭП - до 30%. Чрезвычайно важным преимуществом гетероструктурных GaAs над кремниевыми ФЭП является их способность эффективно преобразовывать 100-1000-кратное концентрированное солнечное излучение в электрическую энергию. В существующих кремниевых СБ эту способность нельзя реализовать из-за существенного снижения величины кпд с увеличением температуры ФЭП. При 30-50-кратном концентрировании излучения величина кпд гетероструктурных GaAs ФЭП составляет 25-27 %. Использовать энергию солнечных элементов можно также как и энергию других источников питания с той разницей, что солнечные элементы не боятся короткого замыкания. Каждый из них предназначен для поддержания определенной силы тока при заданном напряжении. Но в отличие от других источников тока характеристики солнечного элемента зависят от количества падающего на его поверхность света. Важным моментом работы солнечных элементов является их температурный режим. При нагреве элемента на один градус свыше 25°С он теряет в напряжении 0,002 В, т.е. 0,4 %/градус. В яркий солнечный день элементы нагреваются до 60-70°С при потерях каждым 0,07-0,09 В. Это и является основной причиной уменьшения кпд солнечных элементов, что приводит к снижению напряжения, генерируемого элементом. КПД обычного ФЭ в настоящее время колеблется в пределах 10-16 %. Это значит, что элемент размером 100*100 мм$^2$ в стандартных условиях может генерировать 1-1,6 Вт.

В ФЭ и СБ на основе кристаллического кремния (жесткие) кпд 12-20 % в спектре поглощения 0.5-1.0 мкм уменьшается при нагреве (отсчитывая от T = +25°C) на величину 0.45 %/°C; аморфно-кремниевые (гибкие батареи) имеют кпд 5-10 % в спектре 0.2-0.7 мкм; арсенид- галлиевые (жесткие, тяжёлые, дорогие модули) с кпд 10-25 % в спектре 0.2-0.7 мкм сохраняют работоспособность до температуры +150°C; в сульфидно-кадмиевых (тонкоплёноч-

ные – гибкие) кпд 5-10 % в спектре 0.5-0.9 мкм и остается стабильным до Т~ +100°C. Спектры электромагнитных колебаний (длины волн): ультрафиолетовое излучение – 0.005 – 0.4 мкм, видимый свет – 0.36 – 0.75 мкм, инфракрасное излучение – 0.75 – 500 мкм. "Жесткие" батареи – хрупкие, поэтому обычно помещаются в ударопрочный корпус, и их нельзя нагревать более чем до 150-200 $^0$С, иначе произойдут необратимые изменения в контактных слоях и покрытиях. Необходимо использовать герметизирующие оболочки для защиты от коррозии, шунтирования n-p-переходов и потемнения покрытий. Защита особенно актуальна при эксплуатации в химически агрессивной среде городов и в походных условиях; в комплекте надо иметь не только гибкие солнечные батареи, но и обычные кристаллические для возможности получения электропитания при рассеянном освещении, в помещении через «стекло». Срок службы современных солнечных модулей ~ 10 лет. Дольше живут (медленнее «выгорают») кристаллические (жёсткие) батареи. Повышенный кпд имеют ФЭ из кристалла кремния с текстурированной рельефной поверхностью и просветляющим покрытием для минимизации величины светового отражения.

 Гетероструктурные СЭ на основе GaAs имеют более высокий кпд чем кремниевые (монокристаллические и особенно из аморфного кремния), кпд арсенид-галлиевых солнечных батарей доходит до 35-40 %. Максимальная рабочая температура +150 $^0$С в отличие от + 70 $^0$С у кремниевых батарей. Ограничение в температуре полупроводника связано с резким повышением его проводимости. Физическая причина заключается в увеличении темпа тепловой генерации электронно-дырочных пар с ростом температуры, что приводит к рассогласованию внутреннего сопротивления фотопреобразователя с сопротивлением его нагрузки ($R_{внутр.} >> R_{наг}$) и, как следствие, - к снижению напряжения на p-n-переходе (снижение напряжения холостого хода и уменьшение кпд в нагрузке). Тепловой эффект подавляется повышением ширины запрещенной зоны в полупроводнике. Реализация разработанных нанотехнологий позволит осуществить прорыв в решении проблемы увеличения эффективности монокристаллических кремниевых фотопреобразователей для различных энергетических систем [3]. Увеличение токосъема в кремниевых фотопреобразователях является основным фактором повышения их кпд и создания энергоемких солнечных батарей нового поколения. В качестве материала фотопреобразователя солнечной энергии для нового поколения выбран кремний как наиболее изученный, легко доступный и широко распространенный элемент в земной коре. Эти особенности кремния и сравнительно низкая его себестоимость и развитые радиационные технологии несомненно будут определяющим фактором при широкомасштабном производстве солнечной фотоэлектрической энергии на Земле и в условиях космического пространства. Энергетические характеристики СБ определяются полупроводниковым материалом, конструктивными особенностями солнечного элемента, величиной КПД и их количеством в батарее. Наиболее используемыми энергоустановками космических объектов являются СБ на основе кремния ~ 80 %.

# ГЛАВА 1. ФИЗИЧЕСКАЯ МЕТОДОЛОГИЯ РАЗВИТИЯ РАДИАЦИОННЫХ ПРОЦЕССОВ

## 1.1. Каскадные процессы

При прохождении через структуру твердого вещества высокоэнергетическая заряженная частица на атомном уровне создает нарушения, ориентированные вдоль траектории ее движения. Длина макродефекта (трека) различна и изменяется в зависимости от заряда, энергии частицы и атомной структуры вещества. При передаче энергии от заряженной частицы атомам вещества возможны три основных процесса взаимодействия [1]:

– возбуждение электронов атомов вещества и энергетическая релаксация (затухание) в атомах мишени (неупругая часть потери энергии);

– переход кинетической энергии к атомам вещества для прямого смещения атомов из узлов кристаллической решетки (упругая часть потери энергии);

– имплантация чужеродных атомов (стабильных или радиоактивных) внутрь материала (ионное легирование).

В области малых энергий ~100 кэВ преобладают упругие потери, обусловленные рассеянием ионов на атомах твердого тела, а при больших энергиях (несколько десятков МэВ) неупругие процессы преобладают над ядерными и $(dE/dx)_e > (dE/dx)_n$. При неупругом взаимодействии иона с электронами решетки уменьшается его скорость, однако направление его движения остается неизменным. В то же время упругие столкновения иона с атомами не только значительно снижают его энергию, но и могут быть причиной изменения его траектории.

Для высокоэнергетичных ионов с кинетической энергией более 1 МэВ/нуклон интенсивность выделения энергии в электронную подсистему в $10^3$–$10^4$ раза превышает выделение энергии в ядерную подсистему и может составлять несколько МэВ/мкм. Выделение высокой энергии в электронную подсистему усиливает роль электронных возбуждений в генерации дефектов структуры для интенсивного распыления материала, что вызывает локальное плавление, аморфизацию, формирование скрытых треков, генерацию ударных волн и разрушение материала. Релаксация сильных электронных возбуждений является главным параметром, определяющим природу трековых областей. Для большинства металлов и их сплавов при описании трекообразования используется модель «термического клина» при создании в них высоких температур и механизма «кулоновского взрыва» при накоплении заряда в диэлектрических материалах. Для этого вводится величина порогового значения потери энергии частиц, выше которой материал становится чувствительным к образованию ионных треков. Неупругий процесс проходит через две основные стадии. Первая стадия включает диссипацию энергии первичными и вторичными электронами и обмен энергией посредством электрон-электронных столкновений, происходящих за время $10^{-16}$–$10^{-13}$ с. Скорость передачи энергии, обусловленная (e⁻–e⁻) – столкновениями, пропорциональна $\left(\dfrac{dE}{dt}\right)_{e-e} \sim \varepsilon/\tau_{e-e}$, где $\mathcal{E}$ – величина энергии, превышающая энергию электронного возбуждения. Вторая стадия начинается с движения атомов за счет электрон-фононного взаимодействия в течение нескольких дебаевских периодов ($10^{-13}$–$10^{-12}$ с), при этом создается высокая плотность неравновесных дефектов, подвергающихся затем закалке (консервации в решетке после остывания и затвердения). Скорость передачи энергии при электрон–фононных ($e^-$–ph) – столкновениях составляет порядка $\left(\frac{dE}{dt}\right)_{e-ph} \sim \mathcal{E}/\tau_{e-ph}$. Отношение скоростей передачи энергии для двух процессов составляет

$$\left(\frac{dE}{dt}\right)_{e-e} \Big/ \left(\frac{dE}{dt}\right)_{e-ph} \sim \left(\mathcal{E}/kT\right)\left(\tau_{e-ph}/\tau_{e-e}\right) \gg 1. \tag{1}$$

Таким образом, при ($e^-$–$e^-$) – взаимодействии передается большая часть энергии за столкновение $\left(\mathcal{E}/kT \gg 1\right)$, чем при ($e^-$–ph)–взаимодействии.

## 1.2. Релаксационные процессы в металлах с совершенной и несовершенной кристаллической решеткой

Для многих совершенных металлических кристаллов время остывания возбужденной области при передаче энергии электронов кристаллической решетке (электрон-фононное взаимодействие) составляет величину порядка $10^{-12}$ с вследствие малой доли энергии, передаваемой при одном столкновении, а также из-за малого количества столкновений электронов с фононами.

В металлах характерное время остывания возбужденной области за счет электронной теплопроводности определяется из соотношения [1]

$$t_1 = \frac{R^2}{4K_{min}} \approx 10^{-15} c, \tag{2}$$

где R ~ 1 нм – диаметр цилиндрической возбужденной области, $K_{min}$ ~ 1 см²/с – коэффициент электронной термодиффузии.

Следовательно, процесс остывания электронов за счет электронной теплопроводности происходит значительно быстрее, чем передача энергии атомам, и температура, до которой может нагреться решетка, составляет $T_0$ ~100 К. Энергия, переданная электронной подсистеме бомбардирующим ионом, не приводит к распылению атомов структуры, потому что большую часть этой энергии получают электроны проводимости и быстро рассеивают ее по всему большому объему. Поэтому электроны не могут нагреть кристалл до температур, необходимых для распыления, а поскольку в возбужденной области металла нет объемного заряда, то кулоновское расталкивание ионов не реализуется. Таким образом, в идеальном кристалле металла (например, Cu, Al, Ag, Zn, $Cu_3Al$) не должно быть повреждений, вызванных неупругими потерями энергии быстрого иона, что подтверждается отсутствием видимых треков. Тем не менее возможна более интенсивная передача энергии электронов атомам. В несовершенном кристалле вследствие рассеяния электронов на дефектах время электрон-решетчатых релаксаций уменьшается и за время охлаждения возбужденной области по электронам проводимости решетка может успеть приобрести достаточную энергию. В треках тяжелых ионов и осколков деления ядер в металлах Fe, Ti, Zr, Pb, U могут наблюдаться нарушения кристаллической структуры. В этом случае температура в треках превышает температуру плавления материала. За стадией плавления локальных областей следует стадия их сверхбыстрого охлаждения (закалки) с сохранением значительной концентрации дефектов структуры в перекристаллизованной области.

В металлах для объяснения наблюдаемых областей привлекают механизм кластеризации точечных дефектов в результате их термически активируемой миграции вследствие атермических процессов, происходящих в хвосте движущегося быстрого иона. Кроме того наблюдаются эффекты влияния сильных электронных возбуждений на изменение дефектов структуры, созданных за счет упругих столкновений ионов с атомами металла. В чистом никеле и сплаве $Ni_3Fe$ электронные возбуждения вызывают частичный отжиг дефектов, созданных ионами в упругих столкновениях с атомами кристалла, но в Cu, $Cu_3Al$ эффектов ионного облучения не обнаруживается вообще. В других металлах, таких как Ca, Fe, Co, Ti, Zr наблюдается сильное увеличение концентрации создаваемых дефектов, если удельные потери энергии иона на электронные возбуждения превышают определенную пороговую величину. Величина потерь энергии, выше которой образование дефектов резко возрастает, составляют 5 kэВ/нм для сплава FeCo и 40 kэВ/нм для чистого железа [1].

Выделенные состояния релаксационных процессов:

При неупругих взаимодействиях время релаксации области сильного электронного возбуждения в металле с идеальной решеткой происходит за время ($\sim 10^{-15}$ с) на три порядка быстрее, чем передача энергии атомам ($\sim 10^{-12}$ с) при электрон-фононном взаимодействии. Большую часть энергии получают электроны проводимости и быстро рассеивают ее в крупнозернистой структуре по всему объему. Поэтому температура, до которой может быть нагрета решетка с крупнозернистой структурой, незначительная и электроны не могут нагреть кристалл до температуры, необходимой для распыления атомов. В идеальной металлической структуре не возникают повреждения, вызванные неупругими потерями энергии высокоэнергетичных ионов.

В несовершенной кристаллической структуре вследствие рассеяния электронов на дефектах электрон - решетчатые релаксации уменьшаются и за время охлаждения возбужденной области по электронам проводимости решетка может успеть приобрести достаточную энергию. За стадией плавления локальных областей следует стадия их сверхбыстрого охлаждения с сохранением значительных концентраций дефектов структуры в перекристаллизованной области.

Диэлектрики с мелкозернистой структурой имеют большие коэффициенты распыления за счет "кулоновского взрыва" и возбужденные электроны диэлектрических зерен релаксируют медленнее чем в металлах.

В Si–структурах с наноразмерными фрагментами и с повышенной электронной проводимостью возбуждение передается по траектории движения тяжелого осколка ядра и не локализуется в области "горячих" атомов, что не приводит к их испарению. Процесс гидрирования повысит температуру перекристаллизации метастабильной a–Si:H–структуры.

### 1.3. Образование треков в структурах диэлектриков

При ионном облучении диэлектрических материалов [2, 3, 4] электроны вне возбужденной области отсутствуют, поскольку электрическое поле на границе возбужденной области препятствует уходу электронов. При повышенной энергии электронной подсистемы может происходить локальный разогрев микрообласти или даже разрушение облучаемой ионами мишени.

Распыление мелкозернистых диэлектриков определяется отрывом зерна и поэтому возбужденные электроны в них медленнее релаксируют, что может привести к более сильному перегреву и распылению изолированных зерен в виде атомов. Соответствующий перегрев при температуре $T>10^4$ К будет вызывать кипение и химическое разложение среды в треках, а также генерацию мощных механических импульсов. Электронная структура мишени при неупругом распылении проявляется существенно сильнее чем при упругом, поэтому здесь значительно больше доля заряженных распыляемых частиц и в массовом распределении отмечаются кластеры, содержащие сотни и тысячи атомов. В случае магнитных и диэлектрических изоляторов, а также ВТСП–керамики и некоторых проводящих сплавов (Ni-Zr) треки состоят из аморфного материала. Для создания дополнительных возможностей исследования треков в структуре материала часто добавляют в ее состав магнитные элементы. Намагниченная структура в ядерном треке позволяет надежно определять его размеры. Сильные локальные разупорядочения, включая и аморфные структуры, возникающие в ядерных треках, создают парамагнитные центры, снижающие величину насыщенного намагничения, поскольку магнитные свойства наиболее чувствительны к кристаллографическому порядку. Величина радиационно–индуцированного изменения ферромагнитного упорядочения в материале при разупорядочении его решетки определяется методом измерения интенсивности Mossbauer spectra Auger electrons, конвертируемых из K оболочки, а также частично из L и M оболочек атомов $^{57}Fe$ из глубины материала менее 100 нм и методом определения изменения насыщенной намагниченности. Величина насыщенной намагниченности $\mu_s(T,d)$ описывается выражением [4]

$$\mu_s(T,d) = \mu_s(T,0)\exp\left(-\pi R_m^2 d\right), \qquad (3)$$

где $\mu_s(T,0)$ – намагничение при температуре $T$ до облучения, $R_m$ – радиус трека в парамагнитном цилиндре, $d$ – доза облучения.

Экспоненциальная зависимость учитывает состояние трека при высоких дозах облучения. Температурная зависимость $\mu_s(T,d)$ описывается уравнением Brillouin'a. Малое значение отношения величин $\mu_s(T,d)/\mu_s(0,d)$ указывает на отсутствие магнитного порядка внутри трека, т.к. $\mu_s(T,d) < \mu_s(0,0)$. Усредненный диаметр трека ($2R_m$) определяется из выражения

$$\pi R_m^2 d = \ln\left[\frac{\mu_s(T,0)}{\mu_s(T,d)}\right]. \tag{4}$$

Для оценки диаметра трека в диэлектрике SiO$_2$ можно использовать измерения, выполненные в магнитных изоляторах на иттриевом гранате Y$_3$Fe$_5$O$_{12}$, т.к. величина диаметра трека практически не зависит от состава граната [4]. Для иттриевого граната с пороговой потерей энергии ($T_{th}$= 8 кэВ/нм), выше которой эффективность дефектообразования резко возрастает, насыщение достигается при энергии потерь > 22 кэВ/нм [1].

Полые или почти пустые трековые области легко заполняются атомами практически любого сорта за счет процесса трековой миграции атомов из поверхностного слоя. При этом коэффциент диффузии атомов вдоль каналов треков на многие порядки превышает его величину для массивного материала. Учитывая, что диаметр каждого трека составляет ~10 нм, число таких областей на 1 cm$^2$ площади поверхности достигает $10^{12}$ трек/cm$^2$. Плотность упаковки элементов в такой системе наноэлектроники может быть выше на фактор $10^3$, чем предельно допустимая в настоящее время. Для определения корреляции в эффективности образования нарушений структуры и морфологии необходимы прямые наблюдения дефектов. Электронная микроскопия высокого разрешения позволяет провести систематические исследования дефектов, создаваемых ионами [1].

### 1.4. Распыление материала структуры в твердых телах

Энергетическая зависимость коэффициента неупругого распыления $S(E)$ определяется зависимостью неупругих (ионизационных) потерь $(dE/dx)$ от энергии, зарядового состояния бомбардирующих мишень ионов, структурных и термофизических характеристик материала. Для неупругого канала взаимодействия [3, 5–7] энергия ионизационных потерь $(dE/dx)_e$ в веществе хорошо апроксимируется уравнением Bethe–Bloch'а

$$-\left(\frac{dE}{dx}\right)_e = \frac{4\pi q^2 e^4}{mv^2} Z_a N \ln \frac{2mv^2}{j}, \tag{5}$$

где $Z_a$ – атомный номер мишени, $N$ – число атомов в единице объема, $m$ – масса иона, $v$ – скорость иона, $j$ – потенциал ионизации, q – заряд иона.

Наблюдается сильная зависимость $(dE/dx)_e$ от заряда иона, падающего на мишень. Для протона и типичных твердых мишеней $(dE/dx)_e$ не превышает ~100 эВ/нм, в то время как для осколков деления, имеющих скорость ~ $10^9$ cm/c, массу ~100 а.е.м. и заряд $q$ ~20, величина ионизационных потерь энергии $(dE/dx)_e \approx 10^4$ эВ/нм. Ионы тяжелых масс характеризуются большой величиной энергетических потерь. Из функции энергетических потерь следует, что легкие ионы не создают треков. В конкретном варианте усредненные характери-стики, соответствующие легким и тяжелым продуктам деления $^{235}$U, следующие: масса 95±2 и 138.5±1.5 а.е.м, ширина распределения масс 7.5 и 6.6 а.е.м, энергия 99±1 и 67±1 МэВ, начальная скорость 1.41·10$^9$ и 0.97·10$^9$ см/с, заряд ядра $Z_i$ – 38,6 и 53.4 ед. заряда; начальный эффективный заряд $Z_{0i}$ – 22.0 и 16.7 ед. заряда [8]. Характерной особенностью неупругих процессов является большое значение коэффициентов распыления (S) структуры тяжелыми осколками ядер деления [9]. Ионизационные потери осколков больше в начале трека и меньше в конце, а потери энергии осколков на упругие соударения возрастают к концу пробега. Для мелкозернистой структуры (≤10 нм) величина $S(E)$ на несколько порядков больше, чем в крупнозернистых структурах, где осколки деления не оставляют оплавленных областей, поскольку из-за быстрого охлаждения электронного газа решетка не успевает нагреться до температуры плавления [10–12]. Мелкозернистые структуры имеют коэффициент распыления S~10$^3$ атом/осколок [9, 13, 14], в крупнозернистых – S~10 атом/осколок [15, 16]. Таким образом, при неупругом взаимодействии коэффициенты распыления значительны для всех структур материалов в мелкодисперсном состоянии. Крупнозернистые металлы и окислы имеют коэффициенты распыления, не отличающиеся от величины S при упругом распылении. При упругих процессах взаимодействие иона или осколка ядра с решеткой происходит в основном с ядрами атомов или малыми их ассоциациями вследствие каскадного смещения, что приводит к незначительному распылению материала. При этом упругие потери составляют несколько процентов от полных энергетических потерь осколка. Электронная структура мишени при неупругом распылении проявляется существенно сильнее, чем при упругом взаимодействии.

## 1.5. Пороговые величины ионизационных потерь и критерии фрагментации кристаллической структуры

Характерной зависимостью неупругого распыления является его сильная зависимость от структуры мишени. Если осколок деления попадает в фрагмент структуры размером меньше некоторого критического, он распыляется полностью. Чем больше энергия осколка, а точнее $(dE/dx)_e$, тем больше критический размер фрагмента структуры. Если диаметр фрагмента структуры превышает критический размер, коэффициент распыления становится таким же, как у массивной крупнозернистой структуры и по величине сопоставим со значением, которое дает каскадная теория упругого распыления – S ~ 3–4 атом/осколок деления. Значение $(dE/dx)_e$ должно быть достаточно велико, чтобы плотность выделенной энергии в электронной подсистеме превышала некоторую пороговую величину. Это означает, что требуются специальные механизмы распыления и при этом следует обращать внимание на структуру материалов [15]. При малой концентрации смещенных атомов в решетке необходим высокий уровень плотности энергии для плавления структуры. Введение технологического процесса, обеспечивающего высокий уровень дефектообразования и его стабилизацию, позволит значительно снизить величину энергии электронного возбуждения для получения аморфных треков и в монокристаллическом кремнии.

Согласно модели образования треков [17] допускается предположение, что изолированные структуры в зависимости от их размеров и других параметров либо расплавляются полностью, либо не расплавляются совсем, т.е. частичного расплавления не происходит. Причина в том, что плавление при неупругих процессах охватывает область не отдельных атомов, а всю структуру. Последнее указывает на то, что распыление материала происходит за счет ионизационных потерь тормозящихся осколков деления. Минимальные удельные потери энергии на ионизацию определяются из выражения [16]

$$E_{min}/2R = \frac{\frac{4}{3}\pi R^3 N U_0}{2R}, \qquad (6)$$

где $E_{min}$ – минимальная энергия, необходимая для распыления зерна, R-размер зерна сферической формы, N-атомная плотность, $U_0$ – энергия связи атома. При неполном изолировании зерна часть выделенной энергии будет унесена электронами за его пределы, поэтому необходимо ввести коэффициент β-как долю выделенной энергии, которая поглощается в зерне. Тогда

$$E_{min} = \frac{4\pi R^3 N U_0}{3\beta}, \qquad (7)$$

где параметр β зависит от размеров зерна и энергетического спектра δ-электронов. Потери энергии могут определяться электронами проводимости и термоэлектронной эмиссией. Скорость передачи энергии от электронной подсистемы ~$R^3$ (пропорциональна числу атомов в зерне), а скорость потери энергии ~$R^2$ (пропорциональна поверхности зерна). Это накладывает ограничение на минимальный размер зерна. При R< $R_{min}$ энергия, выделенная в зерне, не успевает распространиться среди всех атомов и уйдет за пределы зерна через его поверхность. В реальных образцах с размерами зерен R<$R_{min}$ и R>$R_{max}$ суммарный коэффициент распыления (К) будет определяться размерами площади, которую занимают различные группы зерен

$$K = \left[4\pi N \sum_i f_i R_i^5 \left(1 - 1/n_i^2\right)\right] / 3\sum_i f_i R_i^2, \qquad (8)$$

где $f_i$ –число зерен с радиусом $R_i$ на рассматриваемой площади; величина $\left(1 - 1/n_i^2\right) = P = r_{max}^2/R^2$ – вероятность распыления зерна; $n_i = (dE/dx)_e / (dE/dx)_{e\,min}$ при удельной потери энергии на ионизацию $(dE/dx)_e \geq (dE/dx)_{e\,min}$ и функциональной зависимости $K\left[(dE/dX)_e, R\right]$. Плавление и испарение зерна происходит, если его радиус R≤ $R_{кр}$, определяемый величиной [9]

$$R_{кр} = \left(3\frac{dE}{dx} / 2\pi N U_0\right)^{1/2}. \qquad (9)$$

Для больших размеров зерна вследствие большой длины свободного пробега электронов в электрон-фононных столкновениях по сравнению с длиной свободного пробега в мелкозернистых образцах максимальная температура "термопика" значительно снижается и термический коэффициент распыления становится пренебрежимо малым. В этом случае распыление определяется упругими столкновениями осколка с атомами решетки. Для энергии осколка в несколько десятков МэВ коэффициент распыления уменьшается с увеличением энергии осколка. При этом термический коэффициент распыления мелкозернистой структуры возрастает с увеличением энергии осколка вследствие возрастания ионизационных потерь. При переходе от крупнозернистой области диаметром 5-10 мкм к очень тонким мелкозернистым областям диаметром 5-20 нм резкое изменение коэффициента распыления приходится ориентировочно на размеры зерен 20-30 нм. В работах [7, 18] показано, что для больших размеров фрагментов максимальная величина удельных потерь энергии на образование кластеров в c-Si-структуре составляет 57.2 кэВ/нм при облучении ее фуллеренами $C_{60}$ при их энергии 30-40 МэВ.

1.6. Выводы

Изменение основных свойств c-Si–полупроводников при их облучении частицами, не способными сообщить атомам материала большой энергии, в значительной степени определяется сложными структурными повреждениями типа разупорядоченных областей. Таким образом, при радиационной обработке полупроводниковых кристаллических материалов частицами высоких энергий γ–кванты создают в них только точечные дефекты, а электроны с энергией E ≥ 10 МэВ, нейтроны и ионы создают разупорядоченные структуры за счет каскадных процессов.

# ГЛАВА 2. КОНДЕНСИРОВАННОЕ СОСТОЯНИЕ В КРИСТАЛЛИЧЕСКОЙ СТРУКТУРЕ

2.1. Возможности образования наноструктур в монокристаллическом кремнии

Образование конденсированного состояния в монокристаллическом кремнии связано с эффектом увеличения плотности электронов проводимости в структурах наноразмерной величины. Следует отметить, что электрическая проводимость структуры вдоль траектории движения осколков ядер урана увеличивается. Дислокационная модель строения границ между фрагментами указывает на наличие большого количества оборванных и ненасыщенных спиновых связей у атомов. Время релаксации электрон-фононных взаимодействий ($\iota_{e-ph}$) определяет величину потерь энергии при ее передаче через атомную структуру. Величина $\iota_{e-ph}$ (T) ниже температуры Debye определяет электрическую проводимость σ(T) ее атомной структуры и при T>$T_D$ величина $\iota_{e-ph}$ (T) инвертирует с температурой)

$$\sigma(T) = \frac{N_e e^2 \tau_{e-ph}(T)}{m^*},  \qquad (10)$$

где $N_e$ – плотность электронов проводимости, e– заряд электрона, m*– эффективная масса электрона проводимости. Сброс энергии "горячих" атомов произойдет по наноструктурам макродефекта и атомы достигнут температуры испарения, поскольку $\iota_{e-ph}(T) \sim 1/N_e$. Для получения аморфной фазы в c–Si – структуре время ее остывания должно быть $\sim 10^{-11}$ с.

При взаимодействии высокоэнергетичных электронов с твердым телом создаются различные дефекты, изменяющие свойства материалов. Мерой тормозной способности вещества являются потери энергии быстрой частицы на единице пути – (dE/dx). В соответствии с теорией Бора величина

$$-\frac{dE}{dx} = \frac{2\pi n q^4}{E} \ln \frac{4E}{j}, \qquad (11)$$

где E и q – соответственно энергия и заряд быстрых электронов; n–концентрация атомных электронов в веществе; J – потенциал ионизации вещества (J = 13,6 Z, где Z– эффективный атомный номер вещества).

Траектория движения высокоэнергетичных электронов имеет хаотичный характер и сильно отличается от прямой линии. Полный путь, пройденный таким электроном в поглощающей среде до полной его термализации, определяемый как истинный пробег ($R_c$), рассчитывается по формуле

$$R_c = \int_0^{E_0} \frac{dE}{dE/dx}, \qquad (12)$$

где $E_0$– начальная энергия электронов.

Эмпирическое выражение для экстраполированного пробега ($R_{ex}$) электронов в различных средах, представляющего собой проекцию пути электрона на направление первоначального движения, имеет вид:

$$R_{ex} = 0,4 E^{(1,3-0,1 \lg E)}, \qquad (13)$$

Формула (13) верна для E = (0,01–2,5) MeV. При этом экстраполированный приведенный пробег ($R^1_{ex}$) равен

$$R^1_{ex} = \rho R_{ex}, \qquad (14)$$

где ρ– плотность вещества поглотителя.

Распределение поглощенной дозы по толщине материала поглотителя происходит неравномерно, что является недостатком радиационных технологий, так как в результате облучения материал приобретает неоднородную структуру.

Расчет поглощенной дозы при электронном облучении производится по следующим приближенным формулам:

$$D = \frac{jEt}{e\rho R_{ex}} \quad \text{при } d > R_{ex} \qquad (15)$$

$$D = \frac{jEdt}{e\rho R_{ex}^2} \quad \text{при } d < R_{ex} \qquad (16)$$

Здесь j – плотность тока пучка электронов, t – время облучения, d – толщина материала поглотителя. Механизм прямого дефектообразования частицами (e−, γ, n⁰, ионы, осколки деления ядер $^{238}U$) при формировании наноразмерных a–Si – структур в объеме c–Si – полупроводника реализуется при смещении атомов из узлов кристаллической решетки в междоузельное положение. Выражение для пороговой энергии смещения ($E_d$) первично выбитого атома (ПВА) имеет вид [1]

$$E_d = \frac{2(E_{min} + 2m_0 c^2)}{Mc^2} \cdot E_{min}, \qquad (17)$$

где $m_0$ – масса покоя электрона, M – масса ядра атома облучаемого вещества, c – скорость света. Для Si величина $E_d$ = 12,9 эВ при минимальной энергии бомбардирующих электронов $E_{min}$ ~140 кэВ. При облучении Si электронами с энергией до 10 МэВ в основном происходят неупругие процессы с интенсивным обменом энергией с электронами кристалла (возбуждение электронных оболочек и ионизация атомов, возбуждение колебаний в кристаллической решетке. В случае облучения Si электронами с энергией $E_0 \geq 10$ МэВ упругие столкновения начинают преобладать над неупругими и энергия, передаваемая для смещения атомов, возрастает, но при этом дифференциальное сечение рассеяния уменьшается (см. рис. 1). Для его компенсации необходимо увеличивать дозу облучения с использованием ускорителей с энергией $E_0 > 10$ МэВ и интенсивным пучком электронов. Пороговая энергия образования каскадных процессов смещения атомов составляет $\varepsilon_{dr}$ ~5 кэВ, что соответствует энергии бомбардирующих электронов $\varepsilon_0 \geq 10$ МэВ. Концентрация междоузельных атомов в разупорядоченных областях определяется как $N_{dr} = n_0 \Phi \sigma_{dr}(\varepsilon_0)$, где $n_0 = 5{,}19 \cdot 10^{22}$ см$^{-3}$, Ф – интегральный поток бомбардирующих электронов с энергией $\varepsilon_0$. $\sigma_{dr}(\varepsilon_0) = \int_{\varepsilon_{dr}}^{\varepsilon_{max}} \frac{d\sigma(\varepsilon_0 \varepsilon)}{d\varepsilon} d\varepsilon$ – сечение рассеяния электронов с порогом образования каскадных процессов. Потери энергии электронов при прохождении материала кристалла определяются как $\frac{1}{\rho} \frac{\partial E}{\partial X} = \frac{2 МэВ}{г/см^2}$, где ρ – плотность поглощающего материала. Смещение атомов в каскаде происходит при условии $\varepsilon > \varepsilon_{dr}$, а при величине $\varepsilon < \varepsilon_{dr}$ – атомы выбывают из каскада. При облучении Si-материала нейтронами, α-частицами, протонами и другими тяжелыми ионами наиболее вероятны упругие процессы и энергия, передаваемая первично выбитому атому, больше величины $\varepsilon_{dr}$. Развитые каскадные процессы образуют крупные стабильные структурные нарушения с укрупнением вакансионных кластеров.

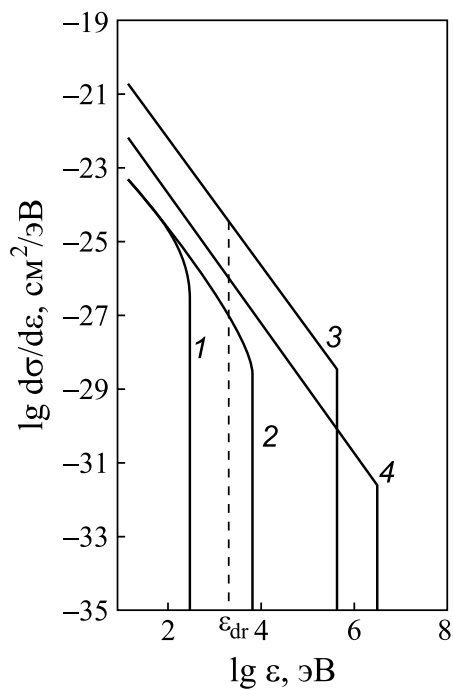

Рис. 1. Дифференциальное поперечное сечение передачи энергии для смещения атомов решетки при бомбардировке кремния электронами с энергией 1 МэВ (1), 10 МэВ (2) и протонами этих же энергий (3 и 4 соответственно) [2].

Высокоэнергетичные γ-кванты

При высокоэнергетическом гамма-облучении ($E_\gamma >$ 0,5 МэВ) образуются точечные нарушения-вакансии и междоузельные атомы. Возникновение смещений в результате непосредственных столкновений γ-квантов с атомами маловероятно. При поглощении γ-квантов происходят Комптон-эффект, образование $e^- – e^+$–пар и фотоэффект. Сечение комптоновского рассеяния $\delta_к \sim Z$–облучаемого материала (Z–определяет число электронов, на которых происходит рассеяние) значительно для фотонов с энергией $E_\gamma = 1–2$ МэВ. Комптон-эффект приводит к появлению во всем объеме кристалла достаточно высокоэнергетичных электронов, способных производить смещения атомов. Сечение образования электрон-позитронных пар $\delta_p \sim Z^2$ возрастает от нуля при $E_\gamma = 2m_oc^2 = 1,02$ МэВ до значительных величин ($\delta_p \cong 10^{-24}$ cm$^2$) для атомов со средним значением Z при энергии гамма-квантов $E_\gamma = 5$ МэВ. Фотоэффект реализуется при $E_\gamma < 1$ МэВ при взаимодействии с K, -L, -M- оболочками атомов вещества. Сечение эмиссии K-электронов ($\delta$) максимально при $E_\gamma$, равной энергии K-электронов, и монотонно уменьшается по закону $\delta \sim E_\gamma$ с возрастанием $E_\gamma$. Равномерное образование точечных дефектов в объеме кристалла определяет основное преимущество облучения γ-квантами перед облучением пучком ускоренных электронов.

При нейтронном облучении Si–материала образуются сравнительно небольшие количества вакансий, способных мигрировать на большие расстояния. Локальная концентрация возникших пар Френкеля может быть достаточно высокой даже с учетом их аннигиляции. Дивакансии, локализованные в небольшом объеме, взаимодействуют между собой и не уда-ляются друг от друга на большие расстояния. Концентрация дивакансий в разупорядоченных областях может составлять ~$10^{20}$ cm$^{-3}$, но эти области даже размером ~$10^2$ Å остаются кристаллическими [1]. Изменение основных свойств Si–полупроводников при облучении их тяжелыми частицами по сравнению с облучением γ-квантами, не способными сообщить атомам материала большой энергии, в значительной степени определяется сложными структурными повреждениями типа разупорядоченных областей. При выборе ускорителя заряженных частиц и гибридного конвертора для постановки эксперимента по разупорядочению структуры материала необходимо учитывать следующие особенности – в развитом электрон-фотонном ливне и тормозном излучении электронная составляющая должна обладать энергией $E_0 \geq 10$ МэВ [3, 4], фотонная составляющая даже высокой энергии создает только точечные дефекты [5, 6]; нейтроны и ионы [7 - 10] создают разупорядоченные структуры за счет каскадных процессов.

2.2. Ядерные процессы

В результате взаимодействия гамма-квантов с ядрами вещества могут происходить различные ядерные реакции, сопровождающиеся выбиванием одной или нескольких частиц (нейтронов, протонов и др.) и возникновением остаточных ядер, более легких, чем ядра исходных материалов (см. табл. 2). Возможность легирования материала с помощью гамма-облучения обусловлена двумя характерными особенностями фотоядерных реакций. Первая заключается в наличии максимума, т.н. гигантского дипольного резонанса, в ходе зависимости величины эффективного сечения (σ) расщепления ядер от энергии γ-квантов ($E_\gamma$). Этот максимум находится при энергии гамма-квантов ~ 15 МэВ, и его резонансная энергия ($E_{res}$) уменьшается с увеличением массового числа ядер A: $E_{res} \approx 75$ A$^{-1/3}$ МэВ, а $E_\gamma \approx 0,7 E_e$. Величина сечения (σ) при $E=E_{res}$ составляет ~50 mbarn для кремния. Такие значения сечения оказываются достаточными для создания требуемых концентраций примесных атомов путем облучения материалов γ-квантами с энергиями до 20 МэВ.

Таблица 2. Перечень фотоядерных реакций (γ, n) и (γ, p) в различных изотопах кремния и порог их возникновения

| Изотоп | Содержание, % | Тип реакции | | | |
|---|---|---|---|---|---|
| | | (γ, n) | | (γ, p) | |
| | | Порог, МэВ | Продукты реакции | Порог, МэВ | Продукты реакции |
| $^{28}$Si | 92,2 | 17,18 | $^{27}$Si→$^{27}$Al $T_{1/2}$=4,2 s | 11,58 | $^{27}$Al |
| $^{29}$Si | 4,7 | 8,48 | $^{28}$Si | 12,33 | $^{28}$Al→$^{28}$Si $T_{1/2}$=2,31 min |
| $^{30}$Si | 3,1 | 10,61 | $^{29}$Si | 13,51 | $^{29}$Al→$^{29}$Si $T_{1/2}$=6,6 min |

Количество примесных атомов алюминия ($N_{Al}$), образующихся в результате облучения кремния тормозными γ-квантами, можно определить с помощью выражения:

$$N_{Al} = J_e \cdot K_{trans} \cdot t, \qquad (18)$$

где $J_e$ - поток ускоренных электронов; $K_{trans}$-коэффициент трансмутации, равный вероятности образования примесного атома, рассчитанной на один падающий электрон, t-время облучения. Величина $K_{trans}$ в значительной степени зависит от энергии электронов, конструкции тормозной мишени и геометрии облучения. В этом случае

$$K_{trans} \sim \sum n_i \cdot A^{\frac{4}{3}}, \qquad (19)$$

где $n_i$ – концентрация исходного изотопа, А – массовое число атома Al.

Как следует из таблицы 2, среди основных продуктов реакции существуют и нестабильные ядра. Поэтому облученные материалы необходимо выдерживать в защитных блоках в течение времени, достаточного для спада наведенной активности до безопасного уровня согласно требованиям Службы дозиметрического контроля. Радиоактивность кристаллов кремния обусловлена, главным образом, ядрами $^{27}Si_{14}$, $^{28}Al_{13}$, $^{29}Al_{13}$ с периодом полураспада не более 10 мин и время спада активности в таких кристаллах будет составлять несколько часов. Отметим, что при радиационном захвате нейтронов (основной канал ядерных реакций при нейтронном легировании) образуются остаточные ядра с массовыми числами (A+1). Именно различия в массовых числах конечных ядер и обусловливают различный состав химических примесей, вводимых в полупроводниковые материалы при их облучении нейтронами и гамма-квантами.

## 2.3. Выводы

Осколки ядер урана характеризуются большой величиной энергетических потерь. Наблюдается сильная зависимость $(dE/dx)_e$ от заряда иона, падающего на мишень. Для неупругого канала рассеяния энергия ионизационных потерь в веществе хорошо аппроксимируется уравнением Bethe–Bloch'a. В результате облучения кремния тормозными гамма-квантами происходит радиационное легирование его структуры фотоядерными реакциями (γ, n), (γ, p) с образованием атомов алюминия (дырочная проводимость в структуре c-Si – фотоэлемента). В продукте (γ, p) - реакции происходит трансмутация атомов $^{29}Al \rightarrow ^{29}Si$ с малым периодом полураспада $T_{1/2}$ = 6,6 мин. Такой ядерный процесс не вызывает компенсацию электронной проводимости дырочной проводимостью в структуре c-Si-фотопреобразователя.

Проблемы солнечной энергетики будут частично решены с привлечением радиационных процессов фотоделения ядер урана и ядерного легирования с процессом трансмутации атомов в структуре полупроводника.

# Глава 3. ФУНКЦИОНАЛЬНЫЕ ОСОБЕННОСТИ МАТЕРИАЛОВ ПЕРСПЕКТИВНЫХ ФОТОПРЕОБРАЗОВАТЕЛЕЙ

## 3.1. Классификация фотопреобразователей

Основные потери солнечного излучения при трансформации его в электрическую энергию определяются фундаментальными ограничениями фотоэффекта в области IR-излучения ($h\nu \leq E_g$) и в области коротковолнового излучения ($h\nu \gg E_g$), а также низкой эффективностью вывода фотогенерированных носителей заряда из объема c-Si-фотопреобразователя [1].

Солнечные батареи относятся к категории приборов, весьма чувствительных к облучению. Поэтому проблема создания радиационно-устойчивых солнечных батарей стоит достаточно остро. На космических объектах используются только СБ $n^+$-p-типа, т.к. радиационные эффекты для p-кремния существенно меньше, чем для n-кремния (тонкий сильнолегированный слой $n^+$-типа с сетчатым или полосковым омическим контактом и протяженный высокоомный участок базы p-типа со сплошным омическим контактом на тыльной стороне). Для реального спектра излучения в космическом пространстве (см. табл. 3) [2], в котором присутствуют как высокоэнергетические, так и низкоэнергетические частицы, деградация коэффициента токосъема будет происходить по всему диапазону длин волн, что приводит к уменьшению выделяемой в нагрузке мощности. Считается, что солнечная батарея сохраняет работоспособность, если мощность в нагрузке уменьшается не более чем на 25 %. При замене диффузионных процессов на дрейфовые можно ослабить влияние диффузионной длины на величину токосъема при облучении фотопреобразователя. Степень деградации СБ под воздействием облучения существенно зависит от параметров орбиты космического аппарата (КА). Особенно опасны радиационные пояса (Van Allen belts ≤ 3000 km). На высоте 1850 км полярной орбиты годовая доза облучения составляет ~ $1,3 \cdot 10^6$ Грей. Энергетический спектр частиц (электроны, протоны) занимает область от нескольких десятков МэВ до энергии ниже 50 кэВ. Для низкоорбитальных КА (770 км) деградация СБ на основе кремния и гетероструктур GaAs-GaAlAs составляет соответственно 15% и 5% в течение 5 лет пребывания КА на. орбите. Для КА на геостационарных орбитах деградация составляет 31% (Si) и 16% (GaAs) в течение 15 лет пребывания на орбите. Для радиационноопасных орбит (7400 км с углом наклона 50°) деградация составляет соответственно 49% (Si) и 22% (GaAs) в течение 5 лет пребывания на орбите. Гетероструктурные ФЭП наряду с повышением эффективности обеспечивают также улучшение радиационной стойкости, что приблизительно в 2 раза увеличивает ресурс работы космических СБ [3].

Таблица 3. Годовые дозы облучения, получаемые поверхностью космического апарата, (Gy)

| Тип излуче- ния | Высота орбиты, км | | | | | | Энергия частиц | Макс. Поток ча- стиц,см$^{-2}$с$^{-1}$ |
| --- | --- | --- | --- | --- | --- | --- | --- | --- |
| | Полярная орбита | | | | Наклон орбиты 30º | | | |
| | 370 | 925 | 1850 | 3700 | 370 | 925 | | |
| Протоны галактичес- кие | $3,5 \cdot 10^{-2}$ | $3,5 \cdot 10^{-2}$ | $4,5 \cdot 10^{-2}$ | $6 \cdot 10^{-2}$ | 0 | 0 | $>10^{13}$ МэВ | ~2,5 |
| Протоны радиацион- ного пояса | $2,6 \cdot 10^{-1}$ | 26 | 13 | $4,3 \cdot 10^{2}$ | $4,3 \cdot 10^{-1}$ | 70 | 0,1-5 МэВ >30 МэВ (3000 км) | $2,8 \cdot 10^{8}$ $8 \cdot 10^{4}$ |
| Электроны радиацион- ного пояса | $1,3 \cdot 10^{3}$ | $1,3 \cdot 10^{5}$ | $1,3 \cdot 10^{6}$ | $7 \cdot 10^{5}$ | $1,7 \cdot 10^{3}$ | $3 \cdot 10^{5}$ | ≤1 МэВ | $2 \cdot 10^{8}$ |
| Протоны (полярное сияние) | 2,6 | 2,6 | 2,6 | 2,6 | 0 | 0 | ~650 кэВ | $10^{5}$ |
| Электроны (полярное сияние) | $4,3 \cdot 10^{4}$ | $4,3 \cdot 10^{4}$ | $4,3 \cdot 10^{4}$ | $4,3 \cdot 10^{4}$ | 0 | 0 | <50 кэВ | $10^{11}$ |
| Протоны (солнечные вспышки) | 26 | 26 | 26 | 26 | 0 | 0 | МэВ- ГэВ | $10^{5}$-$10^{6}$ |
| Суммарное излучение | $4,3 \cdot 10^{4}$ | $1,7 \cdot 10^{5}$ | $1,3 \cdot 10^{6}$ | $7,8 \cdot 10^{5}$ | $1,7 \cdot 10^{3}$ | $3 \cdot 10^{5}$ | | |

Поэтому применение для энергоснабжения космических аппаратов СБ на основе GaAs-гетероструктур дает значительный экономический эффект по сравнению с кремниевыми, несмотря на более высокую их стоимость. Это позволяет снизить расход GaAs полупроводниковых материалов пропорционально степени концентрирования излучения и, следовательно, существенно снизить стоимость солнечной электроэнергии. Их максимальная рабочая температура +150 $^0$С в отличие от + 70 $^0$С – у кремниевых батарей, что увеличивает возможности концентрирования света на ГФП GaAs. ГФП на основе GaAs в значительно меньшей степени чем кремниевые ФЭП подвержены разрушению потоками протонов и электронов высоких энергий вследствие высокого уровня поглощения света в GaAs, а также малых требуемых значений времени жизни и диффузионной длины неосновных носителей. Более того, эксперименты показали [3], что значительная часть радиационных дефектов в ГФП на основе GaAs исчезает после их термообработки (отжига) при температуре как раз порядка 150-180°С. Если ГФП из GaAs будут постоянно работать при температуре ~150 °С, то степень радиационной деградации их будет относительно небольшой на протяжении всего срока активного функционирования станций (особенно это касается космических солнечных энергоустановок, для которых важен малый вес, размер ФЭП и высокий кпд).

В целом можно заключить, что энергетические, массовые и эксплутационные характеристики ГФП на основе GaAs в большей степени соответствуют требованиям для космических условий, чем характеристики кремниевых ФЭП. Но в космических системах до 80% используются сейчас в основном кремниевые фотопреобразователи.

## 3.2. Сравнительные данные ФЭ на основе GaAs и c-Si

Вследствие более высокого уровня поглощения солнечного излучения, определяемого прямыми оптическими переходами в GaAs, высокие кпд ФЭП на их основе могут быть получены при значительно меньшей толщине по сравнению с кремниевым ФЭ. Принципиально достаточно иметь толщину ГФП 5-6 мкм для получения кпд порядка не менее 20 %, тогда как толщина кремниевых элементов не может быть менее 100 мкм без заметного снижения их кпд. Это обстоятельство позволяет рассчитывать на создание лёгких плёночных GaAs ГФП, для производства которых потребуется сравнительно малое количество исходного материала, особенно если в качестве подложки удастся использовать не GaAs, а другой материал, например синтетический сапфир ($Al_2O_3$). Существенное увеличение кпд обеспечивают каскадные ФЭ, изготовленные на основе многослойных гетероструктур с двумя и больше p-n-переходами в материалах с различной шириной запрещенной зоны [3]. Наиболее широко применяются однопереходные ФЭ на основе гетероструктур $GaAs-Al_xGa_{1-x}As$ и тандемные на гетероструктурах GaAs(Ge)-AlGaAs-InGaAs.

В реально действующих структурах с гетеропереходами кпд превышает 30%. Возможность достижения малых начальных значений обратных токов насыщения в p-n-переходах благодаря большой ширине запрещённой зоны позволяет свести к минимуму величину отрицательных температурных градиентов кпд и оптимальной мощности ГФП и, кроме того, существенно расширить область линейной зависимости последней от плотности светового потока. Экспериментальные зависимости кпд ГФП от температуры свидетельствуют о том, что повышение равновесной температуры последних до 150-180 $^0$C не приводит к существенному снижению их кпд и оптимальной удельной мощности. Благодаря устойчивости к высоким температурам арсенид-галлиевые ФЭП позволяют применять к ним концентраторы солнечного излучения. Источники сырья для арсенид - галлиевых фотопреобразователей ограничены ~1 %, процесс его переработки сложный и экологически напряженный. Галлий добывается в основном из бокситов, однако рассматривается также возможность получения его из угольной золы и морской воды. Самые большие запасы галлия содержатся в морской воде, однако его концентрация там весьма невелика, выход при извлечении оценивается величиной всего в 1% и, следовательно, затраты на производство будут вероятно чрезмерно большими. Технология производства ГФП на основе GaAs с использованием методов жидкостной и газовой эпитаксии (ориентированный рост одного монокристалла на поверхности другого) еще не развита до такой степени, как технология производства кремниевых СБ, в результате чего стоимость ГФП сейчас существенно выше на порядки стоимости ФЭП из кремния [3]. Дополнительный резерв экономии галлия связан с использованием в качестве подложки ГФП не GaAs, а синтетический сапфир ($Al_2O_3$). Стоимость ГФП при их массовом производстве на базе усовершенствованной технологии будет также значительно снижена и в целом стоимость системы преобразования энергии на основе ГФП из GaAs может оказаться вполне соизмеримой со стоимостью системы на основе кремния.

Кремний является значительно более доступным и освоенным в производстве материалом, чем арсенид галлия. Кремний широко распространён в природе и запасы исходного сырья для создания ФЭП на его основе практически неисчерпаемы. Технология изготовления кремниевых ФЭП хорошо отработана и непрерывно совершенствуется. Существует реальная перспектива снижения стоимости кремниевых ФЭП на один-два порядка при внедрении новых автоматизированных методов производства, позволяющих, в частности, получать кремниевые ленты, солнечные элементы большой площади и т. п. Здесь играют роль вместе взятые факторы: тарифы, климат, географические широты, способности государства к реальному ценообразованию и долгосрочным инвестициям. Поскольку галлий является весьма дефицитным материалом, что ограничивает возможности производства ГФП на основе GaAs в количествах, необходимых для широкого применения, солнечные элементы поэтому производятся в основном из кремния. Среднее значение кпд в солнечных батареях из монокристаллического кремния сегодня составляет около 12%, хотя и достигает в некоторых случаях ~ 18%. Именно Si СБ в основном можно видеть сегодня на крышах домов разных стран мира.

Очень важна для космических солнечных энергетических систем (СЭС) способность ФЭП не терять кпд при нагревании концентрированным солнечным излучением. Для существующих в настоящее время кремниевых ФЭП повышение температуры свыше 60-70 $^0$C является почти критическим – кпд падает вдвое. Таким образом, в настоящее время трудно отдать явное предпочтение одному из двух рассмотренных полупроводниковых материалов - арсениду галлия или кремнию, и лишь дальнейшее развитие технологии их производства и совершенствование структуры СБ покажет, какой вариант окажется более рациональным для космической и наземной солнечной энергетики [4]. Сейчас пока основным материалом для изготовления солнечных батарей остается кремний.

3.3. Массивы Квантовых нитей в преобразователях энергии солнечного излучения

К настоящему времени большинство коммерческих модулей солнечных элементов основано на кристаллическом кремнии с эффективностью η ≤ 20 % (1-поколение ФЭ) и аморфных тонкопленочных большой площади ФЭ с величиной η ~5-8 % (2-поколение ФЭ). Концепция 3-поколения – это использование нано- и микроструктур (микропроволок). ФЭ с микропроволоками (Vapor-Liquid-Solid - VLS-метод) относятся к классу устройств с особым конструктивным исполнением для получения направленного транспорта НЗ за счет геометрии и материала таких структур. В тонкопленочном ФЭ аморфные слои кремния синтезируются CVD-методом из газов $SiH_4$, $H_2$ и выращиваются на подложке из металлической фольги. Применяется геометрия из набора микроструктур в виде проволок в коаксиальной форме с p-n - переходом. Целью этих исследований является поиск технологических процессов по-вышения эффективности (η) аморфных тонкопленочных ФЭ до величины кристаллических. Особой задачей является повышение эффективности, радиационной стабильности и ресурса работы монокристаллических кремниевых ФЭ с наноструктурами в наземных и космических условиях. Рекомбинационные потери являются основным ограничивающим фактором получения высоких выходных характеристик кремниевых ФЭ. Уменьшение потерь фотогенерируемых НЗ в объеме полупроводниковой структуры в основном достигается за счет улучшения качества кремниевого материала (метод зонной плавки) и уменьшения его толщины с применением оптической системы захвата света. В противном случае время жизни неосновных НЗ ($\tau$) будет снижаться и их диффузионная длина $L = \sqrt{D\tau}$ уменьшаться, где $D = \frac{kT}{q}\mu$ -коэффициент диффузии, μ - подвижность НЗ. При облучении ФЭ в космическом пространстве величина τ = (σ N V)$^{-1/2}$ снижается, рекомбинационные потери НЗ резко возрастают и токосъем уменьшается с потерей кпд (σ - сечение захвата НЗ, N - плотность рекомбинационных центров, V-тепловая скорость НЗ). Для исключения этого фактора вводятся проводящие квантовые структуры в объем полупроводника для вывода неосновных НЗ.

Массивы квантовых нитей в монокристаллическом кремнии с малым удельным сопротивлением формируются двумя радиационными процессами – вследствие частичного разрыва межатомных связей потоком ускоренных электронов и созданием длинных скрытых треков в его структуре за счет процессов ионизации, созданных многозарядными ионами. Степень ионизации структуры кристалла определяет величину зарядовой области при наличии разорванных межатомных связей, снижающих сток заряда из объема полупроводника для возникновения термического пика и кулоновского взрыва с целью увеличения глубины образования проводящих квантовых структур в объеме кремниевого монокристалла. Критерием возникновения скрытого трека считается повышение температуры локальной области кристалла до температуры плавления (аморфная фаза). При высоких величинах выделения энергии (dE/dx ~ Z$^2$) [5] процессы трекообразования, усиленные предварительным дефектообразованием, характерны для каждого иона. При облучении кристалла легкими и тяжелыми осколками ядер урана-238 [6] с величиной заряда $Z_Л$= 38,6; $Z_Т$= 53,4 ед. ионизация структуры вызывает изменение состояния реагирующих дефектов. Ионизация меняет параметры диффузии примесей, что сказывается на глубине их смещения и образования энергетических барьеров для основных НЗ. При малых флюенсах ионов непрерывные треки не формируются, а регистрируются только дефекты и кластеры из-за быстрой эпитаксиальной рекристаллизации вокруг траектории иона (см. рис. 2, 3).

$\tau_{n,p}$

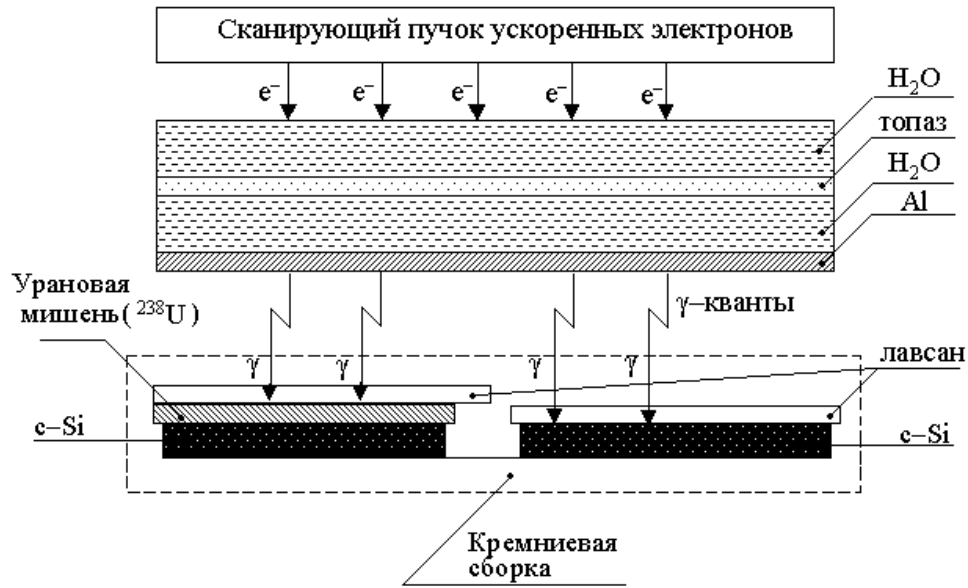

Рис. 2. Схема облучения кремниевой сборки гамма-квантами на линейном ускорителе электронов с энергией 12 МэВ (ускоритель "КУТ-1" ННЦ ХФТИ, Харьков, Украина).

Осколки ядер урана являются малопробежными тяжелыми ионами и глубина их внедрения в материал составляет ~5,2 мг/см², что соответствует для кремния ~22,5 мкм ($\rho_{Si}$=2,3 г/см³). Результаты эксперимента по определению глубины их внедрения в Si-полупроводник без вакансионных кластеров (см. рис. 3) находятся в удовлетворительном согласии с расчетной величиной.

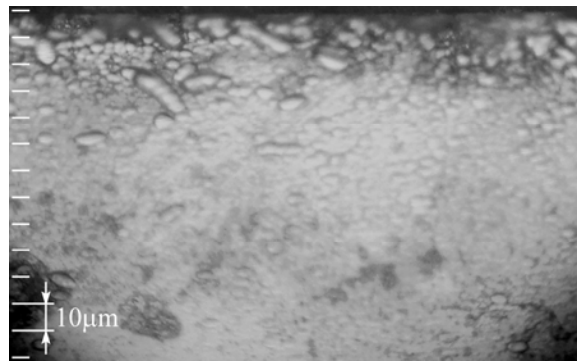

Рис. 3. Микрофотография поперечного среза c-Si-кристалла после облучения его осколками деления ядер урана-238 без дополнительного дефектообразования ускоренными электронами.

Число осколков фотоделения ядер, вводимых в полупроводник, определяется как

$$Y = \frac{1}{2} n_{я} \int_{E_{\gamma пор}}^{E_{\gamma max}} 2\sigma(E_\gamma) f(E_{\gamma max}, E_\gamma) dE_\gamma, \qquad (20)$$

где σ($E_\gamma$) – сечение фотоделения ядер U-238 в зависимости от энергии тормозного γ-кванта, $E_\gamma$, $n_я$ – число ядер U-238 на см², $f(E_{\gamma max}, E_\gamma)$ – трансформированный спектр тормозных γ-квантов, $E_{\gamma\,пор}$=5 МэВ – порог реакции фотоделения ядер U-238, $E_{\gamma\,max}$ – максимальная энер-гия гамма-квантов в спектре тормозного излучения, ½ – коэффициент, учитывающий направление вылета осколков ядер, т.к. при фотоделении урана в области гигантского резонанса угловое распределение легких и тяжелых осколков остается практически изотропным, коэффициент 2 – число осколков деления ядер. Для получения непрерывных треков необходимо повышать флюенс ионов при накоплении достаточно большого количества дефектов структуры и производить предварительное радиационное дефектообразование для срыва процесса рекристаллизации областей около траектории иона. Повышение флюенса осколков производится за счет увеличения сечения фотоделения ядер урана при переходе в область гигантского резонанса с увеличением энергии гамма-квантов (см. рис.4, 5) [6].

Для трансформированного тормозного спектра γ-квантов с максимальной энергией ~ 12 МэВ при поглощенной дозе 0,12 МГр (~8 кГр/час) число осколков ядер урана, вводимое в монокристалл кремния, составляет $Y$=5,4·10$^6$ осколков/см$^2$ час. При необходимости повышения сечения выхода осколков ядер урана до величины 260 mbarn следует увеличить максимальную энергию тормозных γ-квантов до ~ 15 МэВ с максимальным выходом их с конвертора, что соответствует области гигантского резонанса фотоделения ядер U-235, U-238 и увеличению флюенса γ-квантов за счет расширения энергетического интервала от $E_{\gamma\,пор}$ до $E_{\gamma\,max}$ на урановой мишени. Для увеличения сечения фотоделения ядер урана используется ускоритель электронов с энергией 26 МэВ. (ускоритель "ЭПОС" ННЦ ХФТИ, Харьков, Украина).

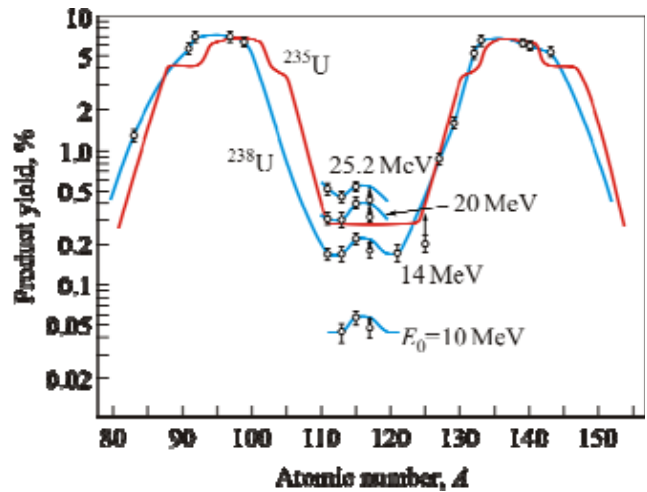

Рис.4. Выходы продуктов фотоделения ядер урана U-235, U-238 от их массового числа А.

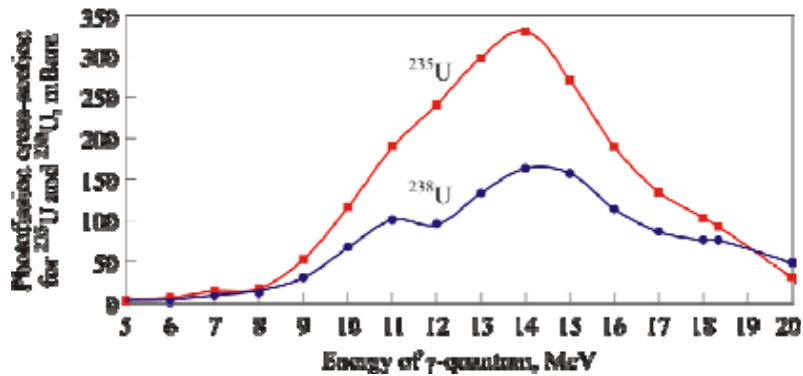

Рис. 5. Энергетическая зависимость сечения фотоделения ядер урана в области гигантского резонанса

Процесс деления определяется минимальным значением кулоновской энергии и энергии поверхностного натяжения делящегося ядра [7]. Деление ядер при энергии их возбуждения, незначительно превышающей высоту барьера деления, является инструментом создания наноразмерных проводящих нитей в легированных c-Si(B,P)-полупроводниках. При выполнении условия $Z^2/A < 49$ деление ядер производится частицами разного типа. Фотоны обычно рассматриваются как наиболее эффективные посредники в передаче ядру объемного возбуждения. Для тяжелых ядер с $Z \geq 90$ ($^{238}_{92}U$) делимость при фотовозбуждении с $E_\gamma \sim 20$ МэВ составляет ~1. Понижение высоты барьера деления в ядрах уранового ряда связано с переполнением их нуклонами. Возбуждение делящейся ядерной системы и образованных осколков снимается испарением нейтронов и γ-квантов и зависит от энергии налетающих частиц. При высоких энергиях возбуждения увеличивается количество испущенных нейтронов, существенно расширяются массовые, энергетические и зарядовые распределения продуктов деления [5, 8, 9]. Начальный эффективный заряд легких и тяжелых продуктов деления составляет $Z_л$=22, $Z_т$=16,7 единиц заряда. Из тяжелых осколков испаряется в среднем больше нейтронов, чем из легких, поэтому энергия гамма-квантов не должна превышать энергию фотопоглощения в области гигантского дипольного резонанса (см. рис. 9). Сечение реакции деления ядер электронами примерно на два порядка меньше, чем в фотоядерных реакциях.

Наличие микро - и наноструктур позволяет решить проблему, возникающую при традиционном способе производства фотоэлементов, которая связана с процессом поглощения солнечного излучения в структуре материала и с выводом носителей заряда из его объема – фотоэлемент большой толщины обеспечивает полное поглощение солнечного излучения, но не обеспечивает эффективный сбор носителей заряда. Поскольку ФЭ обладает низким удельным сопротивлением, то при создании в объеме кристалла глубокозалегающих проводящих квантовых нитей необходима предварительная фрагментация его структуры электронным пучком (разрыв межатомных связей) для сохранения заряда, внесенного осколками ядер урана. Длинные квантовые нити, создающие направленное движение носителей заряда из объема кремниевого полупроводника, при оптимизации толщины p-n-структур обеспечат получение высокой эффективности преобразования солнечного излучения в электрическую энергию в фотопреобразователях нового поколения.

В классической геометрии с планарным p-n-переходом эффективность ФЭ ограничена из-за низкой диффузионной длины неосновных НЗ. В планарной геометрии p-типа $L_n > 1/\alpha$ и $L > 1/\alpha$, где $L_n$-диффузионная длина электронов в p-структуре и любые оптически–генерируемые неосновные НЗ могут достигнуть p-n-перехода без рекомбинации; L-толщина ячейки и при

90 % поглощения первичных фотонов с энергией $h\omega \geq E_g$ составляет 125 мкм; $1/\alpha$ - оптическая толщина материала, зависящая от коэффициента поглощения $\alpha(\lambda)$, интегрированного по всем длинам волн $(\lambda)$. Используя технику улавливания света можно снизить эффективную оптическую толщину материала, что является традиционным приемом в планарной геометрии, но при $L_n < 1/\alpha$ высокоэнергетическая часть солнечного спектра не будет использована в конверсионной эффективности планарных ФЭ.

### 3.4. Выводы

Только кремниевые фотопреобразователи в планарном исполнении с проводящими квантовыми нитями могут представлять практический интерес в разработке ФЭ нового поколения. Фотопреобразователи, изготовленные по такой технологии, по сравнению с типичными объемными кремниевыми фотоэлементами будут обладать большей эффективностью и большим ресурсом эксплуатации для энергетических систем различного назначения, включая и радиационные пояса Земли. Это обещающий кандидат для фотовольтаических устройств нового поколения.

## ГЛАВА 4. КУЛОНОВСКИЙ ВЗРЫВ В МОНОКРИСТАЛЛИЧЕСКОМ КРЕМНИИ

### 4.1. Система управления кластерами

При большом заряде, введенном осколками урана в кристалл, происходит образование кластеров и ионизация их атомов [1, 2]. Атомы кластеров и сложных молекул теряют свои электроны, вызывающие кулоновский взрыв ионного каркаса. Передача энергии от осколков деления валентному электрону атома [3] определяется как $\Delta E_e = 4mE_f / M(1 + 4b^2 m^2 E_f^2 / M^2 Z^2 e^4)$, где $e$, $m$ – заряд и масса электрона, $Z$, $M$, $E_f$ – заряд, масса и энергия осколка, $b$ – прицельное расстояние осколка от атома мишени. Энергия первого ионизационного потенциала (эВ) атомов соответствующих кластеров равна: B-(8,298); Al-(5,986); P-(10,55); Si-(8,151). Площадь поперечного сечения зарядовой области кластера, созданная осколками ядер урана, составляет ~100 Å². Ионизация атомов кластера вызывает его нестабильность, что приводит к высокоэнергетичной диссоциации, делающей каждый атом более положительно заряженным. Атомы области кластера теряют свои электроны и при накоплении положительного заряда ионов происходит "кулоновский взрыв". Если энергия электростатического отталкивания между ионами легированной примеси в кластере превышает энергию связи, то они расталкиваются с большими скоростями. Максимальная энер-гия ионов после "кулоновского взрыва" достигается в случае, когда в процессе взаимодействия с электростатическим полем все электроны покидают кластер. Это возможно, если энергия электронов превышает кулоновский барьер ($U_c = Z \cdot e^2/R$), возникающий в результате их притяжения ионами. Полный заряд кластера равен [3]

$$Z \simeq N = \frac{4\pi}{3} n R^3 \tag{21}$$

где $R$ – радиус кластера, $n$ – концентрация атомов в кластере, $N$ – их полное число. При $R=b$ формируется наноразмерная нить из атомов легированной примеси. Для увеличения радиуса ионизированной области необходимо использовать урановый ряд делящихся атомов.

Минимальное количество атомов (N), необходимое для стабильности кластера с нулевым зарядом, зависит от типа атомов и природы связи между атомами кластера. Большие кластеры легче стабилизируются при высоких ступенях ионизации. Кластеры инертных газов в среднем больше по величине, так как составляющие их атомы имеют замкнутые оболочки и связываются намного более слабыми силами Ван дер Ваальса. Взрыв обнаруживает слоистую структуру, ионы извергаются из кластера последовательно; первая порция ионов имеет высокую кинетическую энергию - она даже выше, чем энергия ионов при кулоновском взрыве малых молекул. Первые быстрые ионы получают энергию от ионов внутренних слоёв кластера и частично от "горячих" электронов. Силы электростатического отталкивания могут превзойти силы притяжения между атомами кластера при введении ионом большой величины заряда в структуру кристалла. При высоких величинах выделения энергии (dE/dx ~ $Z^2$) процессы трекообразования, усиленные предварительным дефектообразованием, присущи каждому многозарядному иону. Сохранение в структуре кристалла введенного заряда позволяет образовывать области кулоновского взрыва на большую глубину.

Состояния поверхности кристалла после облучения его электронами с энергией 12 МэВ дозой $6 \cdot 10^{15}$ e$^-$/см$^2$ при флюенсе пучка 2 мкА/см$^2$, осколками ядер урана и после электрохимического травления приведены на рис. 6, 7.

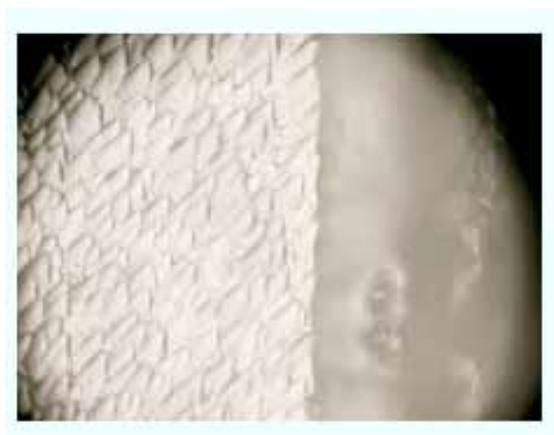

Рис. 6. Разупорядочение атомной структуры электронами на поверхности кремния

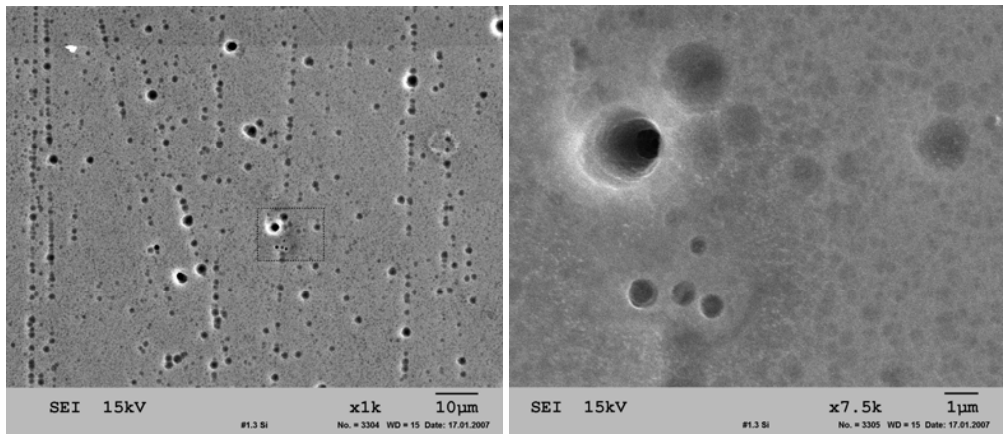

*а*  *б*

Рис. 7. Микрофотография поверхности образца кремния после его облучения осколками ядер урана-238 и электрохимического травления (а, б). На (б) изображен увеличенный фрагмент структуры с фотографии (а), выделенный пунктирной линией.

Для экспериментального обнаружения геометрии скрытых треков облученный образец подвергался кислотному травлению, в результате чего из латентного трека образуется видимый трек в сканирующем электронном микроскопе (см. рис. 8).

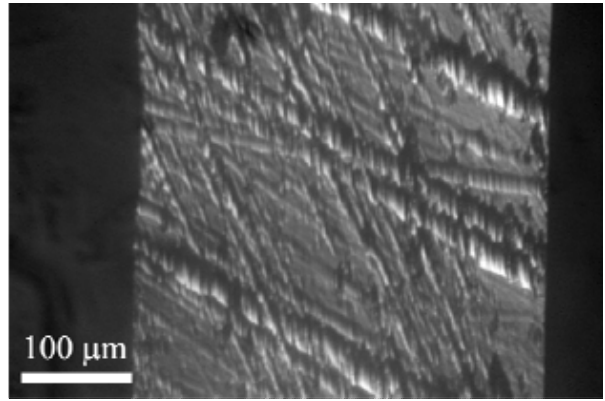

Рис. 8. Микрофотография поперечного среза образца кремния после частичного разрыва атомных связей в его структуре (облучение кристалла дозой 1800 Мрад электронами с энергией 12 МэВ) и после облучения осколками ядер урана-238 при их фотоделении на ускорителе электронов с энергией 26 МэВ.

Поскольку диффузия примесей в аморфной фазе значительно выше, чем в кристаллической структуре, следовательно, кластерные объединения, смещаясь от центра скрытого трека на периферию, создадут наноразмерные нити с увеличенной концентрацией примеси (см. рис. 8). При снижении рекристаллизации на эпитаксиальном слое матрицы область $p^+(n^+)$ будет расширяться в структуру $p(n)$. Краткое представление образования энергетических барьеров для носителей заряда можно предложить в следующей форме

$$C_p(n) \cdot D_p(n))|_{a\text{-Si}} = C_p(n) \cdot D_p(n)|_{c\text{-Si}},\ D_p(n)|_{a\text{-Si}} > D_p(n)|_{c\text{-Si}},$$
$$C_{pi}(ni)|_{a\text{-Si}} < C_p{}^+(n^+)|_{c\text{-Si}},\ \text{т.е. } p(n) < p^+(n^+) \gg pi(ni),$$

где Cp(n), Dp(n) – концентрация и коэффициент диффузии легированных примесей соответственно. При радиационной обработке кремниевых материалов происходит локальная аморфизация кристаллической структуры, что позволяет создавать направленные наноразмерные проводящие нити в объеме кристалла. По окончанию процесса градиентного смещения атомов легирующей примеси остаточное радиационное разупорядочение атомов решетки устраняется отжигом ее структуры. Дополнительный вывод носителей тока из объема кремниевого солнечного элемента вдоль массива наноразмерных проводящих нитей позволяет увеличить КПД солнечных батарей и срок эксплуатации космических объектов.

4.2. Отжиг структуры кремния в кулоновском разряде

Долгоживущая электронно-дырочная плазма кластера нейтрализует кулоновские центры захвата дефектов в разупорядоченной области и таким образом стимулирует диффузию дефектов и примесных атомов на периферию скрытого трека (см. рис. 9)

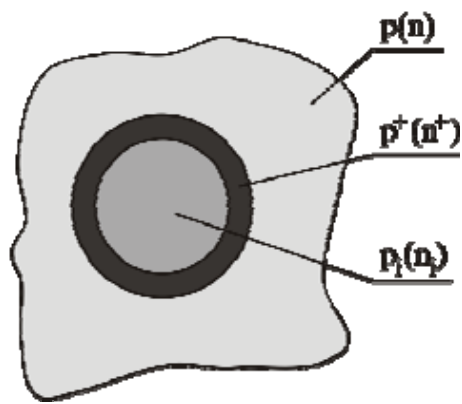

Рис. 9. Образование энергетических барьеров в каналах стока основных носителей заряда

В жидкой фазе скрытого трека [4] коэффициенты диффузии атомов возрастают на 5-6 порядков величины, что обеспечивает условия эффективного создания наноразмерных проводящих нитей из атомов легированной примеси. Только в состоянии жидкой фазы и в присутствии электронно-дырочной плазмы ковалентные связи Si-кристаллов (тетраэдрические полупроводники) настолько ослаблены, что они не стабилизируют фононные моды и эта область аналогична жидкости. Однако эта жидкость отлична от расплавленного Si, т.к. в этом случае энергия, сообщенная кристаллу, первоначально концентрируется в нем в виде электронных возбуждений, а не в виде кинетической энергии колеблющихся атомов решетки, как это имеет место при плавлении - это переход II-рода. Согласно тепловой модели в зависимости от степени нагрева отжигаемого слоя возможны два режима - твердофазной и жидкофазной кристаллизации [5].

Коэффициент диффузии большинства рабочих примесей составляет ~$10^{-10}$ cm$^2$/c в твердой фазе при температуре, близкой к плавлению материала, и ~$10^{-4}$ cm$^2$/c в жидкой фазе. В режиме плавления и кристаллизации происходит сегрегация примесей, описываемая неравновесными коэффициентами распределения. Этот эффект тем сильнее, чем меньше раство-римость данной примеси. Часть примеси оттесняется фронтом кристаллизации к поверхности скрытого трека. В условиях жидкофазной эпитаксиальной кристаллизации процесс растворения примеси в кристалле сводится к захвату атома примеси на границе раздела фаз быстро движущимся фронтом растущего кристалла и к "замуровыванию" его в положении замещения, которое является для него положением с минимумом потенциальной энергии. Вероятность такого захвата определяется соотношением скоростей движения фронта кристаллизации и диффузионного смещения атома примеси в жидкой фазе. Чем больше скорость движения фронта кристаллизации, тем больше атомов примеси захватывается растущим кристаллом. Захваченная примесь уже не имеет возможности для обратной диффузии в расплав или в еще незакристаллизовавшуюся область, поскольку коэффициент ее диффузии в твердой фазе резко (на 5-6 порядков) уменьшается. Таким образом создаются условия для достижения высокой концентрации растворенной примеси на границе скрытого трека. На величину предельно достижимой растворимости примеси влияет фактор концентрационного переохлаждения [5]. Когда концентрация введенной в полупроводник примеси велика, в процессе жидкофазной кристаллизации на границе раздела фаз возникает пик избыточной (над средним уровнем) концентрации примеси в расплаве. Исключительно удобным в технологических целях является использование комбинации облучения кристаллов со стационарным подогревом. При наличии радиационных дефектов в слаборазупорядоченных структурах полупроводника область повышенной концентрации примеси на границе раздела фаз будет размытой, поэтому целесообразно образование жидкофазной области направленным ''кулоновским взрывом'' производить при повышенной температуре кристалла.

Пересыщенные слои являются метастабильными, термодинамически неравновесными и при нагреве релаксируют к более устойчивому состоянию. При этом избыточная примесь покидает замещающее положение в решетке образца, образуя преципитаты или включения вторичной фазы. Предельная температура нагрева, при которой начинается распад пересыщенного раствора, зависит от степени пересыщения и сорта примеси. Эта температура для метастабильных структур в кремнии лежит в пределах 300-950 $^0$C, т.е. достаточно высокая и практически не ограничивает эффективное применение их в полупроводниковой электронике.

Эффекты перераспределения имплантированной примеси неразрывно связаны с диффузией и растворимостью ее в полупроводнике. Поэтому в неравновесных условиях кулоновского разряда процессы диффузии и растворения примеси протекают совершенно иначе, чем при равновесных термических условиях. При переходе от твердофазной кристализации (коэффициенты диффузии малы, время отжига недостаточно для заметной диффузии атомов примеси в твердой фазе) к жидкофазной кристаллизации коэффициенты диффузии примесей изменяются на 5-6 порядков (от $10^{-11}$-$10^{-12}$ до $10^{-4}$ -$10^{-5}$ cm$^2$/c). Кроме того, из-за аномально высоких скоростей движения фронта кристаллизации (~100 cm/c) существенно возрастают пределы растворимости, что оказывает сильное влияние на процессы сегрегации примеси к поверхности скрытого трека. Конечный профиль зависит от индивидуальности атома примеси и плотности энергии в электрическом разряде. Количество сегрегированной примеси к поверхности скрытого трека зависит также и от исходного уровня концентрации легированной примеси и нарушения кристаллической структуры в объеме полупроводника. Скорость и качество кристаллизации зависит также от кристаллографической ориентации монокристалла. Максимальная скорость и наилучшее качество восстановления кристаллической структуры наблюдаются для слоев с ориентацией <100> (дислокационные петли) и худшее – для <111> (двойники). При увеличении плотности энергии в разряде кластера меняется механизм отжига в связи с переходом от твердофазной к жидкофазной кристаллизации. Для режима плавления характерны высокие (> $10^2$ cm/с) скорости кристаллизации, поэтому неравновесные точечные дефекты легко захватываются на границе скрытого трека из-за низкой подвижности их в твердой фазе. Благодаря этому обеспечивается возможность достижения концентрации примесей в замещающих положениях до 1000 раз превышающих равновесные пределы растворимости. Процесс электрической активации внедренной примеси при отжиге в разряде заключается в локализации ее атомов в узлах решетки. При увеличении плотности энергии в разряде происходит увеличение разупорядоченной области с генерацией новых точечных дефектов. Для поддержания режима плавления структуры электронным разрядом кристалл должен иметь повышенную температуру. Отжиг радиационных точечных дефектов управляется диффузионными процессами, определяемыми температурой и временем. В кулоновском разряде при плавлении дефектного слоя точечные дефекты отжигаются практически мгновенно [5], если нет высоких механических напряжений.

4.3. Модификация структуры кремниевых фотопреобразователей

С целью увеличения КПД солнечных элементов в объеме их структур осколками ядер урана создается массив наноразмерных проводящих нитей, по которым из объема полупроводника выводятся носители заряда на электроды внешней электрической цепи фотоэлектропреобразователя (см. рис. 10) [6]. Для увеличения длины проводящих квантовых нитей в c-Si-кристалле необходимо снизить величину потерь энергии по электронам проводимости при плавлении атомной структуры за счет дополнительных разрывов атомных связей электронами во всем объеме кристалла.

Интервал между нитями не должен превышать величину диффузионного смещения ($L_{n,p}$) неосновных носителей заряда:

$$L_{n,p} = \sqrt{D_{n,p}\tau_{n,p}} = \sqrt{\frac{k_B T}{q}\tau_{n,p}}, \quad \tau_{n,p} = \frac{1}{\sigma_{n,p} N_r V_{th}},$$

$N_r = N_c + N_л + N_р$, где $N_r$ – плотность рекомбинационных центров, $N_c$, $N_л$, $N_р$ – плотность собственных, легированных и радиационных центров рекомбинации соответственно, $\tau_{n,p}$ – время жизни носителей заряда, $V_{th}$ – тепловая скорость носителей заряда, $\sigma_{n,p}$ – сечение захвата носителей заряда, $D_{n,p}$ $\mu_{n,p}$ – коэффициент диффузии носителей заряда и их подвижность, $k_B$ – постоянная Больцмана, $T$ – температура кристалла (К), $q$ – заряд электрона (дырки).

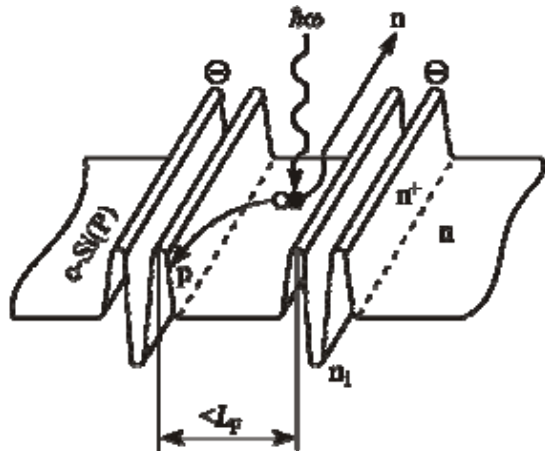

Рис. 10. Схема проводящих наноразмерных нитей для стока носителей заряда в эмиттере фотоэлектропреобразователя (NSC KIPT, Kharkov, Ukraine)

В этом методе смещение атомов легирования приводит к созданию разделительного электрического поля для НЗ и к образованию проводящих квантовых нитей в объеме фотопреобразователя, что обеспечивает увеличение токосъема, повышение эффективности ФЭ и их радиационной устойчивости для использования в космических условиях (см. рис. 10). Канал стока заряда кластера на большую глубину и его направление определяются электрическим полем конденсатора, внутрь которого помещается сборка кристаллов (см. рис. 11). Сканирующий пучок электронов ускорителя заряжает плоский конденсатор, диэлектрической средой которого является кристалл полупроводника.

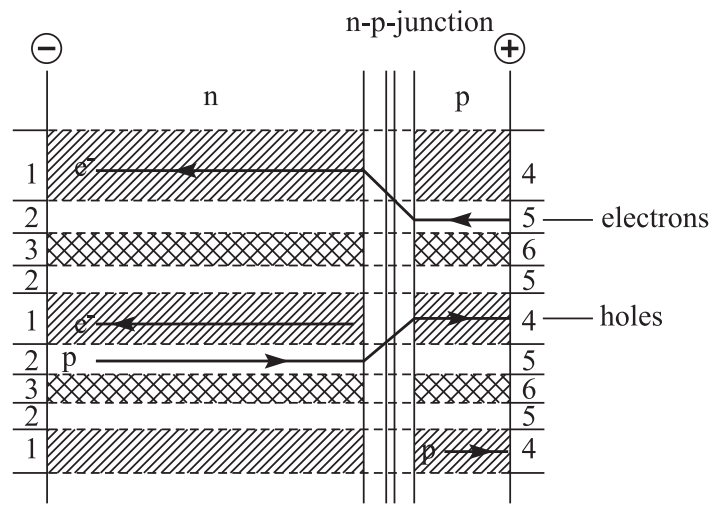

Рис. 11. Движение носителей заряда по квантовым нитям через глубокозалегающий p-n-переход в структуре n >> p (c-Si)-фотопреобразователя 1 – c-Si(n), 2 – квантовые нити с повышенной проводимостью для дырок, 3 – a-Si(ni), 4 – c-Si(p), 5-квантовые нити с повышенной проводимостью для электронов, 6 – a-Si(pi).

Длина проводящей квантовой нити в n-структуре ($L_n$) должна быть максимальной по сравнению с величиной $L_p$ в p-структуре ($L_n \gg L_p$), т.к. структуру c-Si-матрицы сложно бу-дет восстановить до исходного состояния после наработки длинных квантовых нитей. Основными НЗ в n-структурах являются электроны с дрейфовой подвижностью $\mu_n$ ~1500 cm$^2$/В·с в c-Si-полупроводнике, в a-Si-структурах $\mu_n$~10$^{-1}$ cm$^2$/В с как предельное состояние структуры c-Si-матрицы. Поэтому в n-структуре диффузионная длина $L_{dn}$ перемещения электронов к переходу вне квантовых нитей будет ограничена и $L_n \leq L_{dn}$, а дырки (неосновные НЗ) будут беспрепятственно перемещаться от перехода к контактам фронтальной плоскости по проводящим квантовым нитям. В p-структуре величина $L_p$ выбирается минимальной для обеспечения перемещения дырок от перехода к тыльному контакту с пониженной величиной подвижности $\mu_h$ ~500 cm$^2$/В·с в c-Si полупроводнике и очень низкой величиной $\mu_h$ ~10$^{-3}$ cm$^2$/В·с в a-Si-структурах, т.е $L_p \leq L_{dh}$. Электроны (неосновные НЗ) перемещаются от тыльного контакта к переходу по проводящим квантовым нитям в p-структуре. Оптимальное заполнение объема кристалла квантовыми нитями определяется расстоянием ($\ell$) между ними в интервале $\lambda_B < \ell < L$, где $\lambda_B$-длина волны де-Бройля, L-диффузионная длина носителей заряда соответственно.

Массивы квантовых нитей (наноразмерные структуры показаны на рис. 12) являются основным структурным фактором и формируются радиационными процессами, описываемыми моделью термического пика и кулоновского взрыва. Защитное и просветляющее покрытие фронтальной поверхности ФЭ выполняется из алмазоподобной углеродной (DLC) структуры с sp$^2$, sp$^3$ электронной гибридизацией и с минимальным коэффициентом отражения солнечного излучения [7].

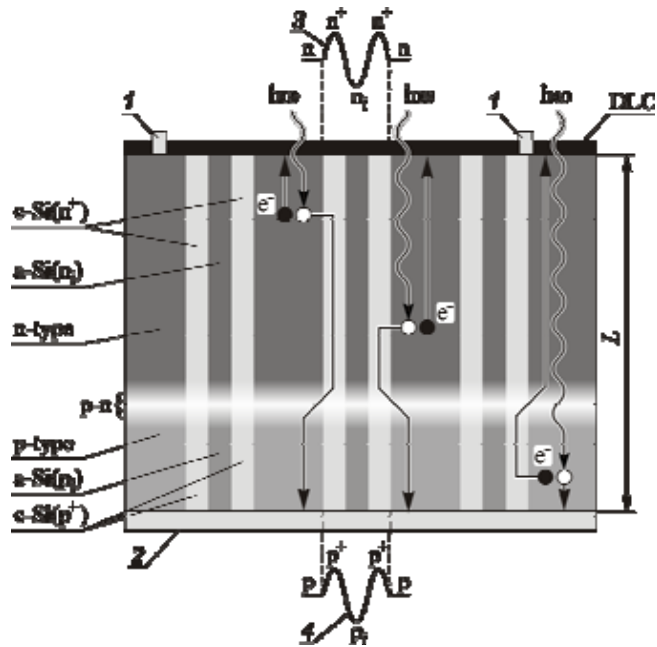

Рис. 12. Схема кремниевого фотопреобразователя в планарном исполнении с проводящими квантовыми нитями (NSC KIPT, Kharkov, Ukraine) 1 –фронтальный контакт, 2 – тыльный контакт, $L$ – толщина ячейки, 3, 4 – форма квантовой нити в n-и p-структурах кремниевого фотопреобразователя.

## 4.4. Стабилизация проводящих нитей в объеме легированного c-Si - полупроводника

Поскольку проводящие нити образуются за счет градиентного смещения атомов легирующей примеси, то радиационные процессы трансмутации Si<Al> не должны снижать уровень металлургического легирования в кремнии. Рассматриваются две схемы радиационных процессов:

1. Каналы фотоядерных (γ, n) и (γ, p)-реакций закрыты и процесс деления ядер урана не обеспечивается (см. табл. 2, раздел 2.2).

В этом режиме наработка нейтронов при фотоядерных реакциях отсутствует, что не создает разупорядочения структуры, влияющей на стабильность проводящих нитей. Основной областью поглощения солнечного излучения является n-структура c-Si(P).

2. При достижении энергии γ-квантов ~ 15 МэВ открываются каналы ядерного легирования за счет фотоядерных реакций, приводящих к увеличению удельного сопротивления в n-структуре c-Si(P) и к снижению удельного сопротивления в p-структуре c-Si(B) – полупроводника.

Для исключения такого фактора основной областью поглощения солнечного излучения является p-структура c-Si(B, Al). Сохранение удельного сопротивления n-структуры производится при ее металлургическом легировании с учетом фактора трансмутации Si<Al> и малого размера области n-структуры. При открытых каналах фотоядерных реакций испущенные нейтроны вызовут дополнительное дефектообразование в c-Si-структуре, что потребует повышения температуры стабилизации и может привести к отжигу проводящих нитей в объеме кристалла. Энергии γ-квантов и нейтронов (~8 МэВ), выделенной при делении ядер урана, еще недостаточно для открытия каналов фотоядерных реакций в кремнии. Дозовые нагрузки на образцах кремния будут минимальными, поскольку облучение кристаллов производится с закрытыми каналами фотоядерных реакций [8].

## 4.5. Выводы

Формирование проводящих квантовых нитей в монокристаллическом кремнии снимает проблему вывода фотогенерированных носителей заряда из объема ФЭ с планарной структурой. В зависимости от процессов облучения и термической обработки c-Si-кристаллов происходит изменение глубины образования проводящих квантовых нитей и уровня восстановления кремниевой структуры вне квантовых нитей до исходного состояния.

## ЗАКЛЮЧЕНИЕ

Основные способы снижения потерь при выводе фотогенерированных носителей заряда из объема c-Si-фотопреобразователя при трансформации энергии солнечного излучения в электрическую энергию следующие:

Образование снижающих аннигиляцию вакансий и междоузельных атомов стабильных кластеров, локальная ионизация слабосвязанных атомов примеси и направленный взрыв за-ряженных областей кластеров на большую глубину – это основные факторы воздействия излучения на структуру легированного c-Si-полупроводника.

Радиационные технологии для преобразования структуры кремниевых солнечных элементов основаны на следующих физических процессах - a) поглощение γ-квантов тяжелыми ядрами для радиационно направленного дефектообразования и ионизации атомов решетки; b) поглощение γ-квантов легкими ядрами для радиационного легирования за счет ядерной трансмутации.

Формирование новых свойств в монокристаллическом кремниевом полупроводнике происходит при пограничном состоянии процессов, когда канал реакции фотопоглощения тяжелыми ядрами максимально открыт, а легкими – еще закрыт.

Для создания скрытых треков в структуре легированного c-Si-полупроводника необходимы большой коэффициент диффузии и увеличенная скорость фронта кристаллизации атомов кремния и легированной примеси, что возможно только в разупорядоченной структуре, находящейся в жидкой фазе.

Экспериментально показана возможность создания с помощью радиационных технологий наноразмерных проводящих нитей для вывода носителей заряда из объема кремниевого полупроводника, что обеспечит повышение КПД преобразования солнечного излучения в электрическую энергию в фотоэлектропреобразователях нового поколения.

# СПИСОК ЛИТЕРАТУРЫ

## К ВВЕДЕНИЮ:

К ГЛАВЕ 1:

К ГЛАВЕ 4:

ПРИЛОЖЕНИЕ

ДОПОЛНЕНИЕ К СПИСКУ ЛИТЕРАТУРЫ

Теоретическая и экспериментальная части решения проблемы повышения эффективности преобразования энергии солнечного излучения в электрическую, изложенные в данной монографии, более подробно рассматриваются в статьях, опубликованных в следующих научных журналах:

1. A.N. Dovbnya, V.P. Yefimov. The Structures for Controlling Charge Carriers in Silicon Photo Cells // Telecommunications and Radio Engineering.- 2008, v. 66, p. 1-18.

Показаны пути создания наноразмерных систем (проводящих нитей) в упорядоченной структуре легированного полупроводника в аспекте современных задач нанотехнологии по усовершенствованию солнечных батарей. Использованы ядерно-физические методы и процессы взаимодействия релятивистских частиц - гамма-квантов, электронов, протонов, нейтронов и осколков ядер тяжелых элементов с атомами вещества, применены радиационные процессы образования кластеров и их направленного кулоновского взрыва.